\documentclass[a4paper,12pt]{article} 
\usepackage[a4paper, left=1in,right=1in,top=1in,bottom=1in]{geometry}
\usepackage{soul}
\usepackage[usenames,dvipsnames]{color}
\usepackage{amsmath, amssymb, slashed, epsf, color, graphicx, latexsym}
\usepackage{epsfig}
\usepackage{graphics}
\usepackage{float}

\usepackage{jheppub}

\newcommand{\be}{\begin{equation}}
\newcommand{\ee}{\end{equation}}
\newcommand{\bea}{\begin{eqnarray}}
\newcommand{\eea}{\end{eqnarray}}
\def\bse{\begin{subequations}}
\def\ese{\end{subequations}}

\def\IZ{\relax\ifmmode\hbox{Z\kern-.4em Z}\else{Z\kern-.4em Z}\fi}

 \def\bsig{{\bar \sigma}}

\def\calE{\mathcal{E}}

\def\presub{\vspace{.5cm} \noindent}

\def\bi{\begin{itemize}} \def\ei{\end{itemize}}

\def\({\left(} \def\){\right)}
\def\[{\left[} \def\]{\right]}
\def\<{\left<} \def\>{\right>}

\title{Distribution of regularized three-body phase-volume}
\author[a,b]{Yogesh Dandekar,}
\author[b]{Barak Kol,}
\affiliation[a]{Department of Physics, Bar-Ilan University, Ramat Gan 5290002, Israel} 
\affiliation[b]{Racah Institute of Physics, Hebrew University, Jerusalem 91904, Israel} 
\emailAdd{yogesh.dandekar@biu.ac.il, barak.kol@mail.huji.ac.il} 

\abstract{The micro-canonical phase-space volume for the three-body problem is a topic of intrinsic interest. Within the flux-based statistical theory, it provides a means to predict the scale of disintegration times for non-hierarchical systems. While the bare phase-volume diverges, arXiv:2205.04294 (Paper I) showed that a regularized version can be defined. 
Building on Paper I, which determined the regularized phase-volume for a given energy $\bar\sigma(E)$, this paper extends the analysis to its distribution over angular momentum, $\bar\sigma(E,L)$. Through analytical integrations, we reduce the problem to a 3d numerical integration, a step up in complexity from the 2d integration required for $\bar\sigma(E)$. We provide regularized phase-volume values for several mass sets across a range of $E$ and $L$, validated through an $L$-integration test. Notably, the values remain positive for all tested parameters, lending further support to the validity of the chosen regularization procedure. For high values of $L$ at fixed masses and $E$, we observe a strong suppression of $\bar\sigma(E,L)$.}

\begin{document}

\maketitle

\newpage
\section{Introduction}

The non-hierarchical Newtonian three-body problem is famously chaotic and is believed to preclude a deterministic solution in closed-form \cite{Poincare_1890}. Instead, one may seek a statistical prediction, which is in fact optimal given the highly random (ergodic) nature of the chaotic component of phase space.  

A statistical theory for this system was formulated in \cite{Monaghan_1976}. Recent important developments include \cite{Stone_Leigh_2019,Ginat_Perets_2020}. However, all of these statistical approaches rely on phase-space volume and on the introduction of the strong interaction radius, a parameter that is not part of the problem data. To remedy that, \cite{flux_based} re-examined the basis for the statistical theory, and presented a different statistical theory, one which relies on the flux of phase-space volume, rather than the volume itself.

The flux-based theory stimulated a novel formulation and a dynamical reduction of the three-body system in \cite{DynRed}. The flux-based theory was tested through simulations in \cite{MTL_2020,simulate,calE}, which provided detailed validation and established the flux-based theory as the most precise statistical theory to date for this system. For recent related work, see \cite{2022MNRAS.517.3838L,Zhang:2023ekp,Trani:2024xwq,Ginat:2024npu,2024MNRAS.tmp.1865G,Klein:2024jbt} and a recent review of Celestial Mechanics \cite{2023ARep...67..742K}.

Within the flux-based theory, the differential decay rate $d\Gamma$ is given by the factorized form \be
d\Gamma(u) = \frac{1}{\bsig}\, \calE(u)\, dF(u)
	\label{bsig_in_flux_based}
\ee
where $u$ is a collective notation for all the outcome parameters over which the decay rate is distributed, $\bsig$ is the regularized phase-volume (short for phase-space volume) of the system, 
$\calE(u)$ is the chaotic emissivity function (equivalently, absorptivity) and $dF(u)$ is the distribution of the asymptotic phase-volume flux. This relation is motivated by an analogy with a particle moving inside a leaky container, where $\bsig$ plays a role analogous to the volume of the container. 

The flux-based theory is the first statistical theory to address the decay rate, thereby going beyond outcome probabilities. The factorization relation \eqref{bsig_in_flux_based} indicates that the scale of the decay rate is set by $\bsig$. Hence, it is essential to evaluate $\bsig$.

The road towards the evaluation of $\bsig$ has begun in \cite{sigma}, here on Paper I. Let us better specify the meaning of $\bsig$ in \eqref{bsig_in_flux_based}. We denote the masses by $m_a \equiv \{m_a\}_{a=1}^3$. First, even though its precise definition involves the chaotic regularized phase-volume $\bsig_\chi$, we shall set out to determine the whole regularized phase-volume, as explained in Paper I. Second, $\bsig=\bsig(E, L; m_a)$, namely $\bsig$ specified by the conserved quantities, the total energy $E$ and the total angular momentum $\vec{L}$. Paper I evaluated $\bsig(E; m_a)$, namely, without specifying $L$, which is easier since the geometry has a higher degree of symmetry before $\vec{L}$ is specified.

Paper I defined the regularization procedure, which was not fully specified in \cite{flux_based}, through subtracting a specific reference integrand: for each spatial configuration, the reference phase space density is given by a hierarchical approximation, where the potential between the binary and the tertiary is replaced by the potential between the binary's center of mass and the tertiary. This makes sense as long as the three-body configuration is indeed hierarchical, and hence, the reference term was supplemented by a condition \be
    E_B \le E/2 ~,
\label{reference_condition}
\ee
where $E_B$ is the relevant binary energy. This condition guarantees that the difference between the bare and reference integrands indeed defines a convergent integral, namely that the divergent-integral singularities of the bare integrand are canceled and that no spurious singularity is added. In addition, this condition has the advantage of being symmetric under a certain involution (reflection), which eventually leads to a fully analytic determination of the $\bsig(E)$ compensator, to be discussed later.

The goal of this paper is to evaluate $\bsig(E,L) \equiv \bsig(E,L;m_a)$, namely the distribution of the regularized phase-volume among the various possible values of $L$. In this sense, it is a continuation of Paper I.

Before we proceed to the evaluation, we discuss why the evaluation of $\bsig(E,L)$ is more challenging than $\bsig(E)$ in Paper I. While the phase space integrand of $\bsig(E)$ is invariant under a general $SO(3)$ rotation, specifying $\vec{L}$ reduces the symmetry to axial $SO(2)$ rotations. Therefore, we shall see that fewer integrations can be performed analytically. While in Paper I, the integration was reduced to a numerical integration over $S^2$, the 2-sphere, here we would be able to reduce it to a numerical integration over $S^3$. The additional dimension of the domain of numerical integration results in a higher computational cost.  

This paper is organized as follows. Section \ref{sec:bare} describes the bare phase-volume, its analytic reduction, singularities, and method of regularization. Section \ref{sec:reg} describes the reduction of the reference and the evaluation of $\bsig$. Section \ref{sec:planar} describes the analogous evaluation for the planar three-body problem. We conclude with a summary and discussion in Section \ref{sec:disc}.

\section{Bare $\sigma(E,L)$ and regularization method} 
\label{sec:bare}

\subsection{Setup of the problem and main goal of this paper} 

The Newtonian three-body problem for point masses is defined by the Hamiltonian
\begin{equation}\label{hm}
    H := \sum_{a=1}^{3} \frac{p_a^2}{2 m_a} - \sum_{a<b} \frac{\alpha_{ab}}{r_{ab}}
\end{equation}
where $m_a$ are the three masses, $p_a$ are their respective momenta, $\alpha_{ab}:=G m_a m_b$ are the potential strength constants, and $r_{ab}:=|\vec r_a-\vec r_b|$ are the separations between the respective pair of bodies.    
The volume of the phase-space for this three-body problem is defined by 	
\begin{equation}\label{bars} 
	\sigma(E,\vec L) := \int \prod_{a=1}^{3} (d^3r_a d^3p_a)\, \delta^{(3)}(\vec R_{cm})\, \delta^{(3)}(\vec P_{cm})\, \delta^{(3)}(\vec J-\vec L)\, \delta(H-E)
\end{equation}
where $E$ is the total energy, $\vec L$ is the total angular momentum, and together they are the  conserved charges of the system. $\vec R_{cm}:=\frac{1}{M}\sum_a m_a\vec r_a$ is center of mass, $\vec P_{cm}:=\sum_a \vec p_a$ is the momentum vector for the center of mass, and $\vec J:=\sum_a \vec r_a \times \vec p_a$. The dimensions of the expression $\sigma(E,L)$ are
\begin{equation}\label{dim3}
	\left[\sigma(E,L)\right] = \frac{M^{3/2}}{E^{5/2}}~ \alpha^3 \equiv [\hbar]^2 T
\end{equation} 
 where $[\hbar]\equiv M L^2 T^{-1}$ denotes the dimensions of action, and $T$ denotes the time dimension. 
 
The phase-volume as defined in \eqref{bars} is divergent, and we need to regularize it in order to extract a finite quantity, rendering it physically meaningful. The regularized phase-volume is denoted by $\bar\sigma(E,L)$. For comparing the theoretical predictions of the Flux-based statistical theory of \cite{flux_based} with the results of numerical simulations (e.g. in \cite{simulate}), we require evaluation of the quantity $\bar\sigma_\chi(E,L)$, which is the regularized chaotic phase-volume. But, as argued in Paper I, $\bar\sigma_\chi(E,L)\simeq \bar\sigma(E,L)$ since chaotic time evolutions are much longer than regular time evolutions. The evaluation of $\bar\sigma(E,L)$ is the main goal of this paper.

\subsection{Analytic integrations and reduction of $\sigma(E,L)$ to $S^3$} \label{cc}

\noindent {\bf Performing integration over momenta}

It turns out to be possible to perform the integration over momenta in expression \eqref{bars} analytically. This was already achieved in Section 3.4 of \cite{flux_based} by introducing auxiliary integration variables. The result is
\begin{equation}\label{smi}
	\sigma(E,L) = 4\pi \left(\frac{M_3}{M}\right)^{3/2} \int_{T_{\text{eff}}\geq0} ~\prod_{a=1}^{3} d^3r_a~ \delta^{(3)}(\vec R_{cm}) \sqrt{\frac{2T_{\text{eff}}}{\det I}}
\end{equation}
where $I^{ij}$ is the moment of inertia tensor for the three-body system defined as
\begin{equation}
		I^{ij} := \sum_{a=1}^{3} m_a \left(r_a^2\delta^{ij}-r_a^i r_a^j\right)
\end{equation}
and $T_{\text{eff}}$ is the effective kinetic energy that accounts for the centrifugal term,
\begin{equation}
    T_{\text{eff}} := E-V-\frac{1}{2} I^{-1}_{ij}L^iL^j
\end{equation}
and $M_3:=\prod_a m_a$, $M:=\sum_a m_a$. 

\presub {\bf Moment of Inertia tensor}

Let us choose the orientation of the Cartesian $x,y,z$ axes so that the three bodies lie in the $x-y$ plane. With this choice, the moment of inertia tensor becomes
\begin{equation} 
	I^{ij} = 
	\begin{bmatrix} 
		I^{xx} & I^{xy} & 0 \\
		I^{xy} & I^{yy} & 0\\
		0 & 0 &  I^{xx}+I^{yy} \\
	\end{bmatrix}
\end{equation}	
where its components are defined by
\begin{equation}\label{mic}
	\begin{split}
		I^{xx} = \sum_a m_a y_a^2 \quad,\quad
		I^{yy} &= \sum_a m_a x_a^2 \quad,\quad
		I^{xy} = -\sum_a m_a x_a y_a \\
		I^{xx}+I^{yy} &= \sum_a m_a r_a^2 = \frac{1}{M}\sum_{a<b} m_a m_b r_{ab}^2
	\end{split}
\end{equation}
Let us introduce the following notation 
\begin{equation}
	\begin{split}
		I & := I^{zz} = I^{xx} + I^{yy} \\
		\det I^{(2)} &:= I^{xx}I^{yy}-\left(I^{xy}\right)^2\\
		\det I &:= \det I^{(2)} I  
	\end{split}
\end{equation}

\presub {\bf Planar reduction}

In the Planar reduction, the coordinate integration in \eqref{smi} over the 3D space is reduced to a coordinate integration over the plane defined by the location of three bodies, and the integration over the orientation of the plane:
\begin{equation}\label{plr}
	\int \prod_{c=1}^{3} d^3r_c~ \delta^{(3)}(\vec R_{cm}) = \frac{1}{2} \left(\int \prod_{c=1}^{3} d^2r_c~ \delta^{(2)}(\vec R_{cm})2A_\Delta\right)\left(\int \sin \theta_n d\theta_n d\phi_n\right)
\end{equation}
The derivation of this result was presented in appendix B of Paper I. $A_\Delta$ is the area of the triangle defined by the three bodies. $\theta_n,\phi_n$ angles define the direction of the normal vector to the plane. 

\presub {\bf Change of coordinates}

As mentioned earlier, we consider the three bodies to lie in the $x-y$ plane, and then perform the planar reduction. Next, we change the 6 Cartesian coordinate variables $x_a, y_a$ to the bi-complex relative position variable $w$ (with 4 real components), and the center of mass coordinates $\vec r_{cm}$ (2 components). The $w$ variable is defined as
\begin{equation}\label{wdef}
	w := (x_1+iy_1) + e^{j\frac{2\pi}{3}} (x_2+iy_2) + e^{-j\frac{2\pi}{3}} (x_3+iy_3)
\end{equation}
where $i$ and $j$ are two independent commuting imaginary units, namely they satisfy the relations $-1=i^2=j^2$ and $i\, j = j\, i$. The $w$ variable was introduced and discussed in the context of three-body systems in \cite{DynRed}. It was also used in Paper I for the reduction of the phase-volume $\sigma(E)$. 

Let us perform another change of variables from $w$ to spherical coordinates over ${\mathbb C}^2$ denoted by $r,\theta,\phi,\psi$ through
\begin{equation}\label{spdef}
	w = r e^{i\psi} \left(\cos \frac{\theta}{2}e_R + e^{-i\phi}\sin \frac{\theta}{2}e_L\right) \quad \text{where,} \quad e_{R,L} = \frac{\sqrt{3}}{2} \left(1\pm ij\right)
\end{equation}
where the angle ranges are $0 \leq \theta \leq \pi,~ 0 \leq \phi \leq 2\pi,~ 0 \leq \psi \leq 2\pi$. 
The variable $r$ defines the overall distance scale. $\theta,\phi$ together define the shape of the triangle created by the three bodies. (The $\theta,\phi$ variables can be thought of as the angles over the ``shape-sphere'' \cite{Lemaitre_1952,Moeckel_Mont_2013,Montgomery2014,DynRed}.) $\psi$ defines the rotation of the triangle within its plane. 

Implementing the two variable changes discussed above, we find 
\begin{equation}\label{coc}
	\begin{split}
		\prod_{c=1}^{3} d^2r_c &= \frac{4}{3}~ d^4 w ~d^2 r_{cm} \\
		&= \frac{3}{4}~ r^3~ \sin \theta ~dr ~d\theta ~d\phi ~d\psi~ d^2 r_{cm} 
	\end{split}
\end{equation}

\presub {\bf Result of analytic reduction of $\sigma(E,L)$ to $S^3$}

Let us summarize the reduction process. First, we perform the planar reduction of \eqref{smi}. Next, we change the Cartesian variables $x_a,y_a$ to the bi-complex $w$ variable and the center of mass, and we perform the integration over the two center of mass variables, which is trivial due to the corresponding $\delta$-function. Next, we change the variables from $w$ to $r,\theta,\phi,\psi$, and perform the integration over $r,\theta_n,\phi_n$ analytically, as described in Appendix \ref{app:A} and in the comments below. As a result, the expression \eqref{smi} is reduced to a numerical integration over the $3$-sphere defined by the angles $\theta,\phi,\psi$, and is given by
\begin{equation} \label{frs}
\begin{split}
    \sigma(E,L) = &\frac{\pi^3}{4\sqrt{2}}  \left(\frac{M_3}{M}\right)\frac{1}{|E|^{5/2}}\int \sin \theta~ d\theta~ d\phi~ d\psi~ \\ &\times\sqrt{\frac{1}{\overline{I}}}(-\bar V)\left[ (3A+B) - 2A \sqrt{\bigg|\frac{A}{B}\bigg|}\theta(-A) \right]\theta(A+B) 
\end{split}
\end{equation} 
where
\begin{equation}
		\begin{split}
			\frac{A}{2|E|L^2} &:= A_1 - A_2 \quad,\quad \frac{B}{2|E|L^2} := A_2 - B_1 \\ A_1 &= \frac{\bar V^2}{2|E|L^2} \quad ,\quad A_2=\frac{\frac{1}{2}\left(\overline{I}+\overline{I}(\psi)\right)}{\overline{\det I^{(2)}}} \quad,\quad B_1 = \frac{1}{\overline I} \\
			\bar I(\psi) &:= (\bar I^{yy}-\bar I^{xx})\cos 2\psi + 2\bar I^{xy}\sin 2\psi 
	\end{split} 
\end{equation}
The overhead bar indicates the removal of $r$-dependence through a multiplication by a suitable power of $r$.\footnote{i.e. $\bar V = r V$, $\bar I = I/r^2$, $\bar I(\psi) = I(\psi)/r^2$, $\overline{\det I^{(2)}}=\det I^{(2)}/r^4$}. Note that $\bar V$ and $\bar I^{ij}$ are functions of $\theta$ and $\phi$, while a dependence on $\psi$ appears only in $\bar I(\psi)$. 

Appendix \ref{apa} presents the derivation of expression \eqref{frs}, and Appendix \ref{rwi} presents the explicit expressions for the various terms in the integrand in $\theta,\phi$ and $\psi$ coordinates. As shown in Appendix \ref{apa},
$$A_1\geq0~,~B_1\geq0~,~A_2\geq0~,~B\geq0$$
which implies that the integrand of \eqref{frs} is positive definite. Finally, Appendix \ref{liv3} validates expression \eqref{frs} by performing the integration over total angular momentum $\vec L$.

\subsection{Divergence of bare phase-volume and the method of regularization}

\noindent {\bf Divergence origin}

The Divergence of the bare phase-volume $\sigma(E,L)$ originates from the hierarchical configurations of the three-body system, where two of the bodies are within a finite distance from each other and the third body is arbitrarily far away.

These hierarchical configurations can be thought of as two decoupled two-body systems. One of them is the binary, which is well separated from the tertiary. The other is the ``hierarchical effective system'' composed of the tertiary and a fictitious center-of-mass body that replaces the binary. For each of the three hierarchical configurations, the three-body potential $V$ can be approximated by the effective potential
\begin{equation}
    V_F:=-\frac{\alpha_B}{r_B}-\frac{\alpha_F}{r_F}
\end{equation}
$-\frac{\alpha_B}{r_B}$ is the binary potential ($V_B$), and $-\frac{\alpha_F}{r_F}$ is the potential for the effective system. We have $\alpha_B:=Gm_am_b$, $\vec r_B:=\vec r_a-\vec r_b$, $\alpha_F:=Gm_s(m_a+m_b)$, $\vec r_F:=\vec r_s- (m_a \vec r_a + m_a \vec r_a)/(m_a+m_b)$, where the $a,b$ indices denote the binary components, and $s$ denotes the tertiary. \footnote{It is understood that $V_B$ and $V_F$ depend on the identity of the tertiary.}

In the limit $r_F\rightarrow \infty$, the potential $V_F$ is dominated by the binary potential $V_B$. Since the integrand is almost constant and the integration over $r_F$ is unbounded, the integral diverges in this limit.
\footnote{Other two special configurations could also potentially be (but are not) sources of divergence. First, when all three bodies are far away from each other. Second, when the distance between two bodies is arbitrarily small, and the third body is at a finite distance away. However, it can be argued that these special configurations do not lead to a divergence. See Section 2.1 of Paper I for the related discussion of divergence of bare phase-volume $\sigma(E)$.}

\newpage
\presub {\bf Regularization}

We regularize the bare phase-volume by subtracting a reference term.  This term eliminates the divergence, and makes the resulting regularized phase-volume a finite and meaningful physical quantity. The subtraction of the reference term from the bare phase-volume is made in the following manner
\begin{equation}
    \bar\sigma(E,L) := \sigma(E,L) - \sigma_{\text{ref}}(E,L) \quad \text{at the level of integrand}
\end{equation}
where the integration is over three-sphere $S^3$ (discussed in section \ref{cc}). This method of subtraction avoids the occurrence of infinite quantities, and the integration over $S^3$ remains finite. Note that since both $\sigma(E,L)$ and $\sigma_{\text{ref}}(E,L)$ are individually positive, their subtraction, i.e. $\bar\sigma(E,L)$ can be either positive or negative. A positive result can be interpreted as giving rise to a finite decay time for the three-body system. In contrast, a vanishing or negative result can be interpreted as the breakdown of the statistical theory.

\presub {\bf Reference phase-volume}

The reference phase-volume is defined as 
\begin{equation}\label{rpv}
    \sigma_{\text{ref}}(E,L):= \sum_{a=1}^3 \int_{D_s} \prod_{c=1}^{3} (d^3r_c d^3p_c) \delta^{(3)}(\vec R_{cm})\delta^{(3)}(\vec P_{cm}) \delta^{(3)}(\vec J-\vec L)\delta(H_{F,s}-E)
\end{equation}
where the Hamiltonian $H_{F,s}:=T+V_{F,s}$ is obtained by replacing the three-body potential $V$ of the bare phase-volume with the effective potential $V_{F,s}$ for tertiary ($s$). The integration domain $D_s$ is defined as the set of all points in phase-space that satisfy the condition $E_B\leq E/2$. This condition (same as chosen in Paper I) is physically motivated (see Section 2.2 of Paper I for the motivation) and helps to regularize the bare phase-volume to produce physically reasonable results presented later in section \ref{res3}

Note that the choice for the integration domain $D_s$ is not unique. Another choice of integration domain would change the final values for the regularized phase-volume. However, the physical motivation associated with our choice and the resulting values of the regularized phase-volume support our choice of the integration domain. \\\\
{\bf Implementation of regularization}

Practically, we choose to implement the regularization 
through a different and simpler scheme. This simpler scheme is defined by \eqref{rpv}, except we choose the integration domain $D'_s$ given by condition $u_B\geq1/2$, where $u_B:=V_B/|E|$. It can be shown that $E_B\leq E/2\implies u_B\geq1/2$. We find the answers for the desired regularization scheme $E_B\leq E/2$ by calculating the compensator defined as
\begin{equation}\label{sref}
    \Delta \sigma (E,L):=\sigma_{\text{ref}'}(E,L)-\sigma_{\text{ref}}(E,L) \quad \text{at the integrand level}
\end{equation}
where $\sigma_{\text{ref}}(E,L)$ is the reference term with integration domain $D_s$ given by the $E_B\leq E/2$ condition, and $\sigma_{\text{ref}'}(E,L)$ is the reference term with integration domain $D'_s$ given by the $u_B\geq 1/2$ condition. Later we will evaluate \eqref{sref}.

\section{Regularized $\bar\sigma(E,L)$}
\label{sec:reg}

\subsection{Analytic integrations and reduction of $\sigma_{\text{ref}',s}(E,L)$ to $S^3$}\label{anre}

{\bf Reference phase-volume}

As discussed in the previous section, we implement the regularization by finding the regularized phase-volume for the simpler scheme and then adding the compensator term. 
The reference phase-volume $\sigma_{\text{ref}',s}$ is obtained by the replacement $V\rightarrow V_{F}$ in the expression for the bare phase-volume and imposing the $u_B\geq\frac{1}{2}$ condition  \footnote{This condition can be rewritten in terms of $r,\theta,\phi$ coordinates as $r\leq\frac{2\bar \alpha_B}{\rho_B(
 \theta,\phi
 )|E|}$.}. After performing the reduction of the reference phase-volume similar to the bare phase-volume presented in Appendix \ref{apa}, we get
\begin{equation}\label{sre}
\begin{split}
   		\sigma_{\text{ref}',s}(E,L) &= 3\sqrt{2} \pi^2 \left(\frac{M_3}{M}\right) \int \sin \theta d\theta d\phi d\psi \sqrt{\frac{1}{\overline I}}  \int \sin \theta_n d\theta_n \int dr  \\&\quad\quad  \times r \sqrt{-|E|r^2-{\bar V_{F}}r-\frac{1}{2}{\bar I}^{-1}_{ij}L^iL^j}^+ 
\end{split}
\end{equation}
where the integration domain is defined by the positivity condition on the integrand and the $u_B\geq\frac{1}{2}$ condition. Notice that expression \eqref{sre} is same as expression \eqref{sr5} under the replacement $V\rightarrow V_F$. \\\\
{\bf Further integration over coordinate variables}

Let us proceed to perform analytic integrations over the $r$ and $\theta_n$ variables (and reduce the reference term to a numerical integration over $S^3$ defined by $\theta,\phi,\psi$ coordinates). For convenience, let us introduce some notations (relevant only for the current section \ref{anre}) as follows \footnote{Actually $f(r,\theta_n)$ is a function of $r,\theta_n,\theta,\phi,\psi$ variables. But we only specify $r,\theta_n$ dependence explicitly to make the notation simpler.}
\begin{equation}\label{rwe}
\begin{split}
    f(r,\theta_n) &:= r \sqrt{-|E|r^2-{\bar V_{F}}r-\frac{1}{2}{\bar I}^{-1}_{ij}L^iL^j}^+ \\
     G_s(\theta,\phi,\psi) &:= \int \sin \theta_n d\theta_n \int dr~ f(r,\theta_n) \\
    \implies \sigma_{\text{ref}',s}(E,L) &= 3\sqrt{2} \pi^2 \left(\frac{M_3}{M}\right) \int \sin \theta d\theta d\phi d\psi~\sqrt{\frac{1}{\overline I}}~ G_s(\theta,\phi,\psi)
\end{split}
\end{equation}

Now we reduce \eqref{sre} further. Let us define $\tau:=\cos\theta_n$. Performing similar calculation steps as in Appendix \ref{apa}, we get
\begin{equation}\label{idec}
	\begin{split}
		-\frac{1}{2}\bar I^{-1}_{ij} L^iL^j &= C+D\tau^2 \\
		\text{where,}\quad\quad  C &= -\frac{L^2}{2} \frac{\frac{1}{2}\left(\overline{I}+\overline{I}(\psi)\right)}{\overline{\det I^{(2)}}} \\
		D &= \frac{L^2}{2} \left(\frac{\frac{1}{2}\left(\overline{I}+\overline{I}(\psi)\right)}{\overline{\det I^{(2)}}} - \frac{1}{\overline I}\right)
	\end{split}
\end{equation} 
It can be shown that
\begin{equation}
	C \leq 0 ~,~D \geq 0 
\end{equation}
Substituting \eqref{idec} in \eqref{rwe}, we get
\begin{equation}\label{fred}
	f(r,\tau) = r \sqrt{-|E|r^2-{\bar V_{F}}r+C+D\tau^2}^+
\end{equation}
Positivity of the argument of the square root implies 
\begin{equation}
	\tau^2 - \frac{|E|}{D}r^2-\frac{\bar V_{F}}{D} r + \frac{C}{D} \geq 0
\end{equation}
which we rewrite as
\begin{equation}\label{hyre}
	\begin{split}
		&\tau^2-p\bar r^2+q\geq0 \\
		\text{where,}~~ p &= \frac{|E|}{D} ~,~ 
		q = \frac{1}{D}\left(C+\frac{\bar V_{F}^2}{4|E|}\right) ~,~
		\bar r= r+\frac{\bar V_{F}}{2|E|} 
	\end{split}	
\end{equation}
Let us consider the $r-\tau$ plane, with $r$ axis horizontal and $\tau$ axis vertical. For the case of equality in the expression \eqref{hyre}, we get two hyperbola curves. When $q\geq0$ the hyperbolas are `horizontal' and when $-1\leq q\leq0$ the hyperbolas are `vertical'. Let
\begin{equation}
	G_s(\theta,\phi,\psi) = G_s^{(1)}(\theta,\phi,\psi)+G_s^{(2)}(\theta,\phi,\psi)
\label{ref_reduced_schem2}
\end{equation}
where $G_s= G_s^{(1)}$ when $q\geq0$ and $G_s= G_s^{(2)}$ when $-1\leq q\leq0$. Let us find the $\tau,r$ integration domain for these two cases.\\\\

\begin{figure}[h!]
	\includegraphics[scale=0.25]{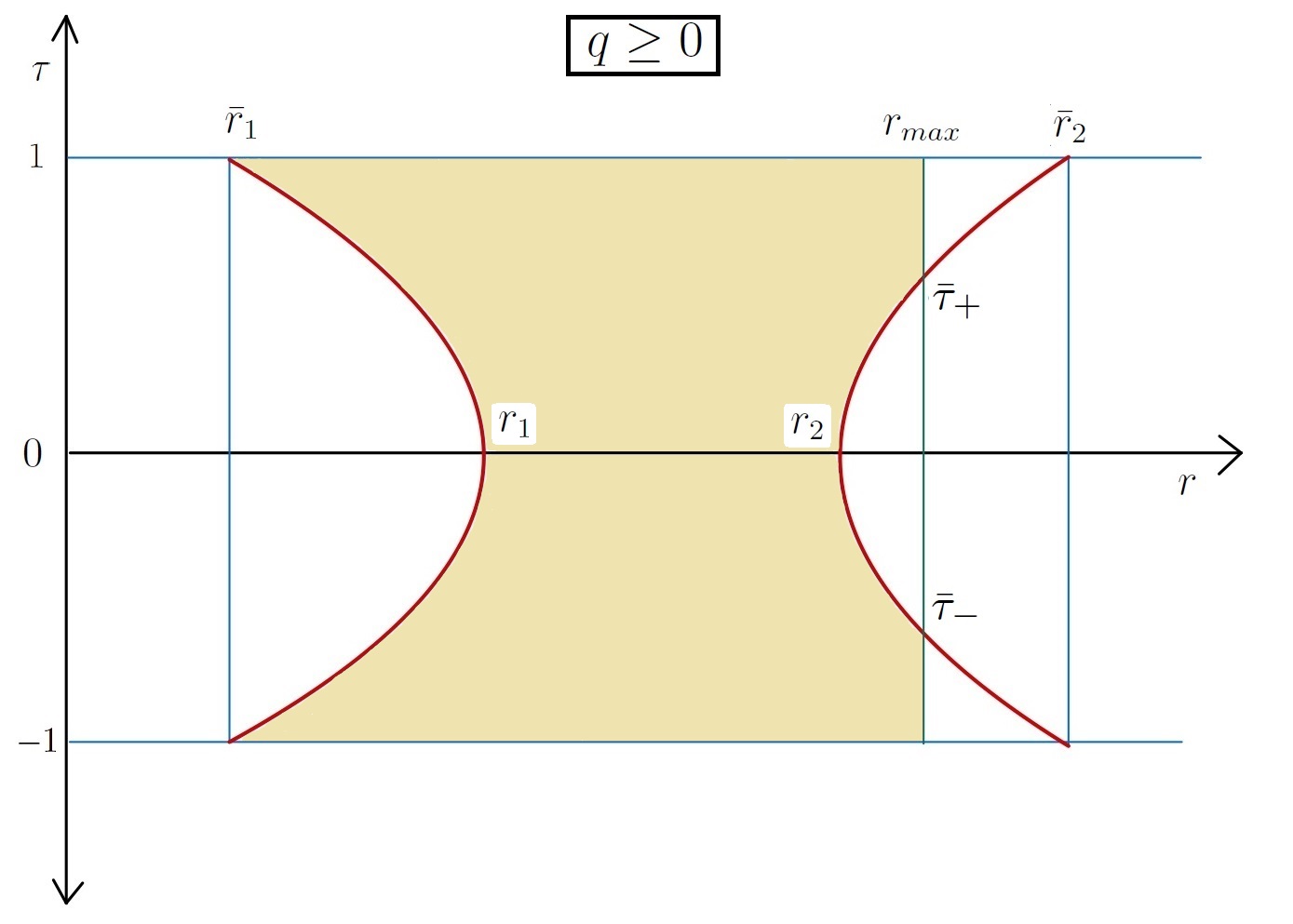}
	\caption{ Integration domain (shaded by light yellow) for $q\geq 0$. Horizontal hyperbolas are shown in red. Second case from expression \eqref{qgz} is shown as a representative.}
	\label{qgf}
\end{figure}

{\bf Case 1: $q\geq0$}, see figure \ref{qgf}. We perform integration over the $r$ variable first, and then over the $\tau$ variable. The $r$ integration is defined by the intersection of two conditions: $f_-(\tau)\leq r\leq f_+(\tau)$ and $0\leq r\leq \frac{-2\bar V_B}{|E|}$, where $f_\pm(\tau)\equiv \pm\sqrt{\frac{\tau^2+q}{p}}-\frac{\bar V_{F}}{2|E|}$. Let us define the following quantities:
\begin{equation}
	\begin{split}
		r_1 &\equiv -\sqrt{\frac{q}{p}} -\frac{\bar V_{F}}{2|E|} ~~,~~ 
		r_2 \equiv \sqrt{\frac{q}{p}} -\frac{\bar V_{F}}{2|E|} \\
		\bar r_1 &\equiv -\sqrt{\frac{q+1}{p}} -\frac{\bar V_{F}}{2|E|} ~~,~~ 
		\bar r_2 \equiv \sqrt{\frac{q+1}{p}} -\frac{\bar V_{F}}{2|E|} \\
		r_{max} &\equiv \frac{-2\bar V_B}{|E|} ~~,~~ \bar \tau_{\pm} \equiv \pm\sqrt{p\left(r_{max}+\frac{\bar V_{F}}{2|E|}\right)^2-q} 
	\end{split}
\end{equation}
where, the left (right) hyperbola cuts the $\tau=\pm1$ lines at $r=$ $\bar r_1$ ($\bar r_2$) and cuts $\tau=0$ line at $r=$ $r_1$ ($r_2$). In case it happens, then the $r=r_{max}$ line cuts either the left or right hyperbola at $\tau=\bar \tau_{+}$ and $\tau=\bar \tau_{-}$. After substituting values of $p,q$, we see that $\bar r_1\geq0$ for all possible values of parameters.

According to the value of $r_{max}$, we get four terms  in the expression of $G_s^{(1)}$
\begin{equation}\label{qgz}
	\begin{split}
		&G_s^{(1)}(\theta,\phi,\psi) = \theta(q)\Bigg[\theta(r_{max}-\bar r_2) \left(\int_{-1}^{1} d\tau \int_{f_-(\tau)}^{f_+(\tau)} dr f(r,\tau) \right) \\&+  \theta(\bar r_2-r_{max})\theta(r_{max}-r_2) \left(\int_{\bar \tau_-}^{\bar \tau_+} d\tau \int_{f_-(\tau)}^{f_+(\tau)} dr f(r,\tau) + 2 \int_{\bar \tau_+}^{1} d\tau \int_{f_-(\tau)}^{r_{max}} dr f(r,\tau) \right) \\ &+ \theta(r_{max}-r_1)\theta(r_2-r_{max}) \left(\int_{-1}^{1} d\tau \int_{f_-(\tau)}^{r_{max}} dr f(r,\tau)\right) \\ &+\theta(r_1-r_{max})\theta(r_{max}-\bar r_1) \left( 2 \int_{\bar \tau_+}^{1} d\tau \int_{f_-(\tau)}^{r_{max}} dr f(r,\tau) \right) \Bigg]  		
	\end{split}
\end{equation}

\begin{figure}[h!]
	\includegraphics[scale=0.25]{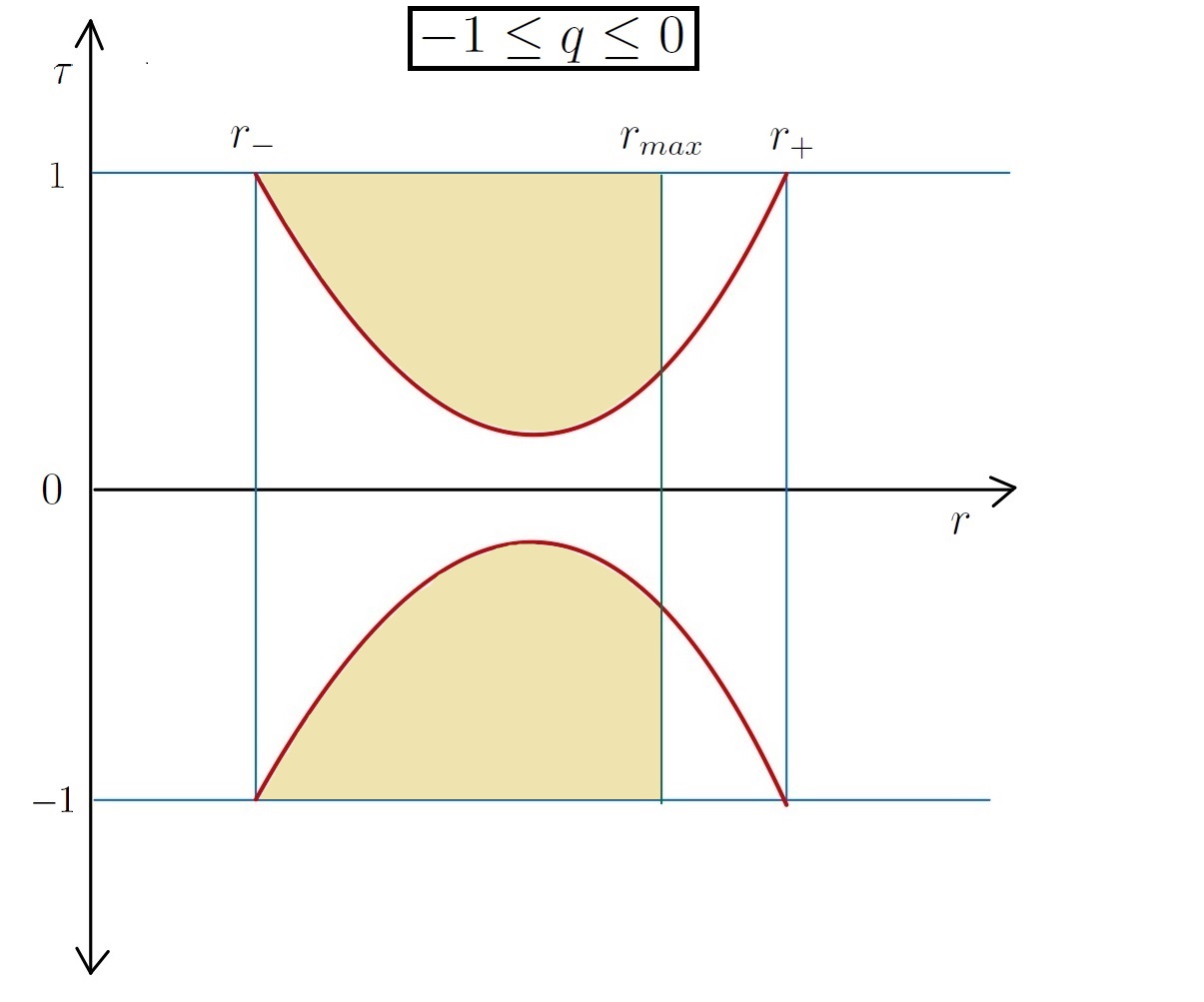}
	\caption{Integration domain (shaded by light yellow) for $-1\leq q\leq 0$. Vertical hyperbolas are shown in red. Second case from expression \eqref{qlz} is shown as a representative. }
	\label{qlf}
\end{figure}

{\bf Case 2: $-1\leq q\leq0$}, see figure \ref{qlf}. We perform integration over the $\tau$ variable first, and then over the $r$ variable. The $\tau$ integration is between $g_+(r)\leq \tau\leq1$ and $-1\leq \tau\leq g_-(r)$, where $g_\pm(r)\equiv \pm\sqrt{p\left(r+\frac{\bar V_{F}}{2|E|}\right)^2-q}$. The hyperbolas intersect the $\tau=\pm 1$ lines at $r_\pm \equiv \pm\sqrt{\frac{q+1}{p}}-\frac{\bar V_{F}}{2|E|}$. There is an overall restriction of $0\leq r\leq \frac{-2\bar V_B}{|E|}$ on the integration domain. We find that $r_-\geq0$ for all values of parameters. Since $f(r,\tau)$ is an even function of $\tau$, we can write 
\begin{equation}
	\int_{g_+(r)}^1 d\tau f(r,\tau) + \int_{-1}^{g_-(r)} d\tau f(r,\tau) = 2 \int_{g_+(r)}^1 d\tau f(r,\tau)
\end{equation}

According to the value of $r_{max}$, we get two terms in the expression of $G_s^{(2)}$
\begin{equation}\label{qlz}
\begin{split}
    G_s^{(2)}(\theta,\phi,\psi) = &\theta(-q)\theta(q+1)\Bigg[\theta(r_{max}-r_+) \left(2\int_{r_-}^{r_+} dr \int_{g_+(r)}^1 d\tau f(r,\tau) \right) \\&+ \theta(r_+-r_{max})\theta(r_{max}-r_-) \left(2\int_{r_-}^{r_{max}} dr \int_{g_+(r)}^1 d\tau f(r,\tau) \right)  \Bigg]  
\end{split}
\end{equation}
{\bf Final expression:}
The final expression for $G_s(\theta,\phi,\psi)$ is given by substituting\eqref{qgz} and \eqref{qlz} into \eqref{ref_reduced_schem2}. By substituting $G_s(\theta,\phi,\psi)$ in \eqref{rwe} we get the expression for reference phase-volume. Hence, the reference phase-volume is a piecewise function constructed from 6 cases. It turns out to be possible to perform analytic integrations over the $\tau,r$ variables, and the results are presented in \href{https://drive.google.com/drive/folders/1JzuRxzQnGMjZlZOVWE1clLu3ZDtC-5m9?usp=sharing}{Supplementary Material}
Hence, the reference phase-volume is reduced to a numerical integration over $S^3$, similar to the bare phase-volume.

\newpage
\subsection{Compensator $\Delta\sigma(E,L)$}
{\bf Explicit form of the compensator}

 For hierarchical configurations, there are two decoupled two-body systems defined by the binary separation and momentum vectors $\vec r_B,\vec p_B$ and by $\vec r_F,\vec p_F$ for the effective system. Let $E_B,\vec L_B$ denote the energy and angular momentum of the binary system, and let $E_F=E-E_B,\vec L_F=\vec L-\vec L_B$ denote the energy and angular momentum of the effective system. The compensator can be expressed in terms of effective variables as
\begin{equation}\label{com3}
	\begin{split}
		\Delta\sigma_{s}(E,L) = &\int dE_B \int d^3 L_B \int d^3 r_B d^3 p_B \delta(H_B-E_B)\delta^{(3)}(\vec J_B-\vec L_B)  \times \\& \times\int d^3 r_F d^3 p_F \delta(H_F-(E-E_B))\delta^{(3)}(\vec J_F-(\vec L-\vec L_B)) \\
		\text{with,}\quad \quad & \\
		H_B = \frac{p_B^2}{2\mu_B}& - \frac{\alpha_B}{r_B} ~,~ H_F = \frac{p_F^2}{2\mu_F} - \frac{\alpha_F}{r_F} ~,~ \vec J_B = \vec r_B \times \vec p_B ~,~ \vec J_F = \vec r_F \times \vec p_F 		
	\end{split}
\end{equation}
where the domain of integration of \eqref{com3} is defined by two conditions: $E_B\geq E/2$ and $u_B\geq1/2$.  \\\\

\newpage
\noindent {\bf Reduction of compensator}

Performing the integration over the momenta $\vec p_B,\vec p_F$ we find \footnote{The integration over momenta can be performed by introducing variables conjugate to the momenta, analogous to the method used in section 3.4 of \cite{flux_based}.}
\begin{equation}\label{ebm}
	\begin{split}
		&\Delta\sigma_{s}(E,L) = 2 \left(\frac{M_3}{M}\right)^{1/2}\int dE_B \int d^3 L_B \int \frac{d^3 r_B d^3 r_F}{r_B^2r_F^2} ~\times \\& \times \delta (L_B^{\parallel})\delta (L_F^{\parallel})\left(E_B + \frac{\alpha_B}{r_B}-\frac{(L^\perp_B)^2}{2\mu_B r_B^2} \right)^{-1/2}  \left(E_F + \frac{\alpha_F}{r_F}-\frac{{(L_F^\perp)}^2}{2\mu_F r_F^2}\right)^{-1/2}
	\end{split} 
\end{equation} 
where
\begin{equation}
	E_F=E-E_B ~~,~~ \vec L_F=\vec L-\vec L_B ~~,~~ L_B^\parallel=\hat r_B\cdot \vec L_B~~,~~\vec L_B^\perp=\vec L_B - L_B^\parallel \hat r_B
\end{equation}
Let us use spherical coordinates $r_B,\theta_B,\phi_B$ for $\vec r_B$. Choose the orientation of this spherical coordinate system so that $\vec L_B$ lies along the direction of $\theta_B=0$. Perform integration over $\theta_B,\phi_B$ analytically. After performing the analogous steps for $\vec r_F$, we get
\begin{equation}
	\begin{split}
		\Delta\sigma_{s}(E,L) &= (8\pi^2)  \left(\frac{M_3}{M}\right)^{1/2}\int dr_B dr_F \int dE_B \int d^3 L_B~\times \\& \times \frac{1}{|L_B||L_F|} \left(E_B + \frac{\alpha_B}{r_B}-\frac{L_B^2}{2\mu_B r_B^2} \right)_+^{-1/2}  \left(E_F + \frac{\alpha_F}{r_F}-\frac{L_F^2}{2\mu_F r_F^2}\right)_+^{-1/2}
	\end{split}
\end{equation}
After changing the variables $r_B,r_F$ to $u_B,u_F$ defined as
\begin{equation}\label{udef}
    u_B:= \frac{\alpha_B}{r_B |E|}~~,~~u_F:= \frac{\alpha_F}{r_F |E|}
\end{equation}
we find
\begin{equation}\label{refeb}
	\begin{split}
		\Delta\sigma_{s}(E,L) &= (8\pi^2)  \left(\frac{M_3}{M}\right)^{1/2} \frac{(\alpha_B\alpha_F)}{|E|^2} \int \frac{d u_B d u_F}{u_B^2u_F^2}  \int dE_B \int d^3 L_B~\times \\& \times \frac{1}{|L_B||L_F|} \left(E_B + \frac{\alpha_B}{r_B}-\frac{L_B^2}{2\mu_B r_B^2} \right)_+^{-1/2}  \left(E_F + \frac{\alpha_F}{r_F}-\frac{L_F^2}{2\mu_F r_F^2}\right)_+^{-1/2}
	\end{split}
\end{equation}
Next, use spherical coordinates $L_B,\theta_L,\phi_L$ for $\vec L_B$. Choose the orientation of this spherical coordinate system so that $\vec L$ lies along the direction of $\theta_L=0$. The integration over $\phi_L$ becomes trivial (and gives a multiplicative factor $2\pi$). Defining $t:=\cos \theta_L$, we find
\begin{equation}\label{comp3}
	\begin{split}
		\Delta&\sigma_{s}(E,L) 
		= (16\pi^3) \left(\frac{M_3}{M}\right)^{1/2} \frac{(\alpha_B\alpha_F)}{|E|^2}\int dE_B \int d L_B\int dt \int \frac{d u_B}{u_B^2}\int \frac{d u_F}{u_F^2}  \\& \times\frac{L_B}{L_F} \left(E_B + u_B|E| - \frac{L_B^2|E|^2}{2\mu_B\alpha_B^2}u_B^2 \right)_+^{-\frac{1}{2}}  \left(E_F +u_F|E| - \frac{L_F^2|E|^2}{2\mu_F\alpha_F^2}u_F^2\right)_+^{-\frac{1}{2}}
	\end{split} 
\end{equation}
where
\begin{equation}
	L_F=\sqrt{L^2+L_B^2-2LL_Bt}
\end{equation}
Performing the integration over $u_F$ and $u_B$ analytically, we find
\begin{equation}
    \begin{split}
        &\Delta\sigma_{s}(E,L) 
		=(16\pi^3) \left(\frac{M_3}{M}\right)^{1/2} \frac{(\alpha_B\alpha_F)}{|E|^2} \int_{{\cal R}_1}  dE_B~ dL_B~dt~\times\\& \quad \quad \quad \quad \times \frac{L_B}{L_F} \left[\frac{\pi}{2}\frac{|E|}{(|E|+E_B)^{3/2}}\right] 
		\left[\frac{\pi}{2}\frac{|E|}{(-E_B)^{3/2}}\right]
          \\&+ (16\pi^3) \left(\frac{M_3}{M}\right)^{1/2} \frac{(\alpha_B\alpha_F)}{|E|^2}  \int_{{\cal R}_2}  dE_B~ dL_B~dt~\frac{L_B}{L_F} \left[\frac{\pi}{2}\frac{|E|}{(|E|+E_B)^{3/2}}\right]\times \\&\times \Bigg\{
		\Bigg[\frac{2}{E_B}\sqrt{\frac{|E|}{2}+E_B-\frac{|E|^2L_B^2}{8k_B}}+\frac{|E|}{4(E_B)^{3/2}}\log\left(\frac{-1+\frac{|E|+4E_B}{4\sqrt{E_B}\sqrt{\frac{|E|}{2}+E_B-\frac{|E|^2L_B^2}{8k_B}}}}{1+\frac{|E|+4E_B}{4\sqrt{E_B}\sqrt{\frac{|E|}{2}+E_B-\frac{|E|^2L_B^2}{8k_B}}}}\right)\Bigg]\theta(E_B)
  \\& +\Bigg[\frac{\pi}{4}\frac{|E|}{(-E_B)^{3/2}}-\frac{2}{(-E_B)}\sqrt{\frac{|E|}{2}+E_B-\frac{|E|^2L_B^2}{8k_B}}\\&-\frac{|E|}{2(-E_B)^{3/2}}\arctan\left(\frac{|E|+4E_B}{4\sqrt{(-E_B)}\sqrt{\frac{|E|}{2}+E_B-\frac{|E|^2L_B^2}{8k_B}}}\right)\Bigg]\theta(-E_B) \Bigg\}
    \end{split}
    \label{compensator_reduced}
\end{equation}
where ${\cal R}_1$ and ${\cal R}_2$ are the two regions of integration given by
\begin{equation}
	\begin{split}
		{\cal R}_1 = &\left(\frac{E}{2}\leq E_B\leq \frac{E}{4}\right) \text{and} \left(E_B\leq E+ \frac{k_F}{2\left( L^2+L_B^2-2LL_Bt \right)}\right) \\ &\text{and} \left( \frac{-k_B}{2L_B^2}\leq E_B\leq \frac{E}{2}+\frac{E^2L_B^2}{8k_B}\right) \text{and}~ (-1\leq t\leq 1) \\
		{\cal R}_2 = &\left(\frac{E}{2}\leq E_B\leq \infty\right) \text{and} \left(E_B\leq E+ \frac{k_F}{2\left( L^2+L_B^2-2LL_Bt \right)}\right) \\ &\text{and} \left(  \frac{E}{2}+\frac{E^2L_B^2}{8k_B}\leq E_B\right)  \text{and}~ (-1\leq t\leq 1)
	\end{split} 	
\end{equation}
The remaining integrations over the $L_B,E_B,t$ variables are performed numerically. Note that the integral converges, even though the integrand diverges in the limit $L_B\rightarrow L,t\rightarrow1$. The compensator satisfies the $L$ integration validation (performed numerically), i.e. it is consistent with compensator $\Delta\sigma(E)$ in Paper I defined for the regularized phase-volume $\bar\sigma(E)$.

\newpage
\subsection{Evaluation of $\bar\sigma(E,L)$}\label{res3}

{\bf Numerical integration and addition of Compensator}

The bare phase-volume $\sigma(E,L)$ has three singularities which lie on the $\theta=\pi/2$ equator at $\phi=0,2\pi/3,4\pi/3$. These points correspond to triangles in which two vertices coincide.  
We find that the integrand of the regularized phase-volume is regular at the three coincidence singularities, and so the integral is indeed finite as expected\footnote{Note that the regularity of the integrand at the location of the singularities is incidental. For the case of $\sigma(E)$ in Paper I, the integral converges near the location of singularities, yet the integrand is singular.}. We perform the numerical integration (using Mathematica) to find the regularized phase-volume using the $u_B\geq1/2$ scheme. While performing numerical integration, we exclude a small region near the coincidence singularities to get reliable answers. While taking this excision to zero, we observe that the answers remain stable. We then numerically evaluate and add the compensator to find the answers for the desired $E_B\leq E/2$ scheme.\\\\
{\bf Normalization}

We define the normalized and dimensionless phase-volume $\hat\sigma(E,L)$ as
\begin{equation}
    \begin{split}
        \hat\sigma(E,L) &:= \frac{1}{(2\pi)^6}\left(\frac{M}{M_3}\right)^{3/2}\frac{|E|^4}{\sigma_0} (\bar L^2)^{3/2} \bar \sigma(E,L) \\ \sigma_0 &:= \sum_{s=1}^3 (\alpha_{B,s}\alpha_{F,s})^3  \quad \text{and} \quad \bar L^2 := \frac{1}{2|E|}\frac{\sum_{s=1}^3 k_s}{3} 
    \end{split}
\end{equation}
where $k_s:=\mu_B \alpha_B^2$ ($\mu_B$ is the reduced mass for the binary). This normalization is a simple modification of the normalization used in Paper I. $\bar L^2$ denotes a typical measure for the (square of) total angular momentum. \\\\
{\bf $L$ dependence and $L$ integration validation}

\begin{figure}
	\includegraphics[scale=1]{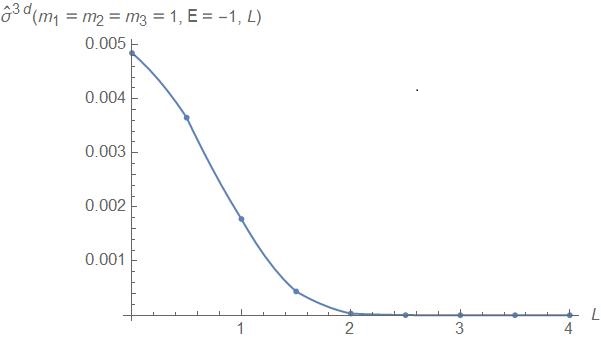}
	\caption{$L$ dependence of $\hat\sigma(E,L)$ for $m_1=m_2=m_3=1$ and $|E|=1$.}
	\label{svl3}
\end{figure}

We set $m_1=m_2=m_3=1$, $|E|=1$, and observe the $L$ dependence of $\hat\sigma(E,L)$ using scatterplots. We also construct an interpolation curve for guidance. In Fig. \ref{svl3}, we overlap the scatterplot and interpolation curve for $\hat\sigma(E,L)$. For the range of values of $L$ considered, we see that $\hat\sigma(E,L)$ is a monotonic function of $L$, remains positive, and drops sharply as $L$ is increased.  We numerically performed the $L$-integration validation for $\bar\sigma(E,L)$ using the interpolation function, and we found that it is satisfied (within 3\% relative error). \\\\
{\bf $\bar\sigma(E,L)$ for 8 Mass sets of MKTL \cite{simulate}}

In \cite{simulate}, the authors performed one million simulations for the three-body dynamics for each of the 8 chosen mass sets and plotted the distributions of disintegration times. They considered initial conditions so that the total energy and total angular momentum are given by
\begin{equation}
    E = -\frac{m_a m_b}{2a} - \frac{m_s(m_a+m_b)}{d} ~,~ L = m_a m_b \sqrt{\frac{a}{m_a+m_b}} 
\end{equation}
where $d=100$AU, and $a=5$AU are the distances that specify the initial condition of a bound binary and tertiary falling towards each other. $m_a,m_b$ are the masses of the initial binary, and $m_s$ is the mass of the tertiary. They use units where $G=1$, mass is measured in units of $M_\odot$, and distance is measured in units of AU. For these 8 mass sets, we present the corresponding regularized phase-volumes in table \ref{sb3}.

\begin{figure}
	\includegraphics[scale=0.65]{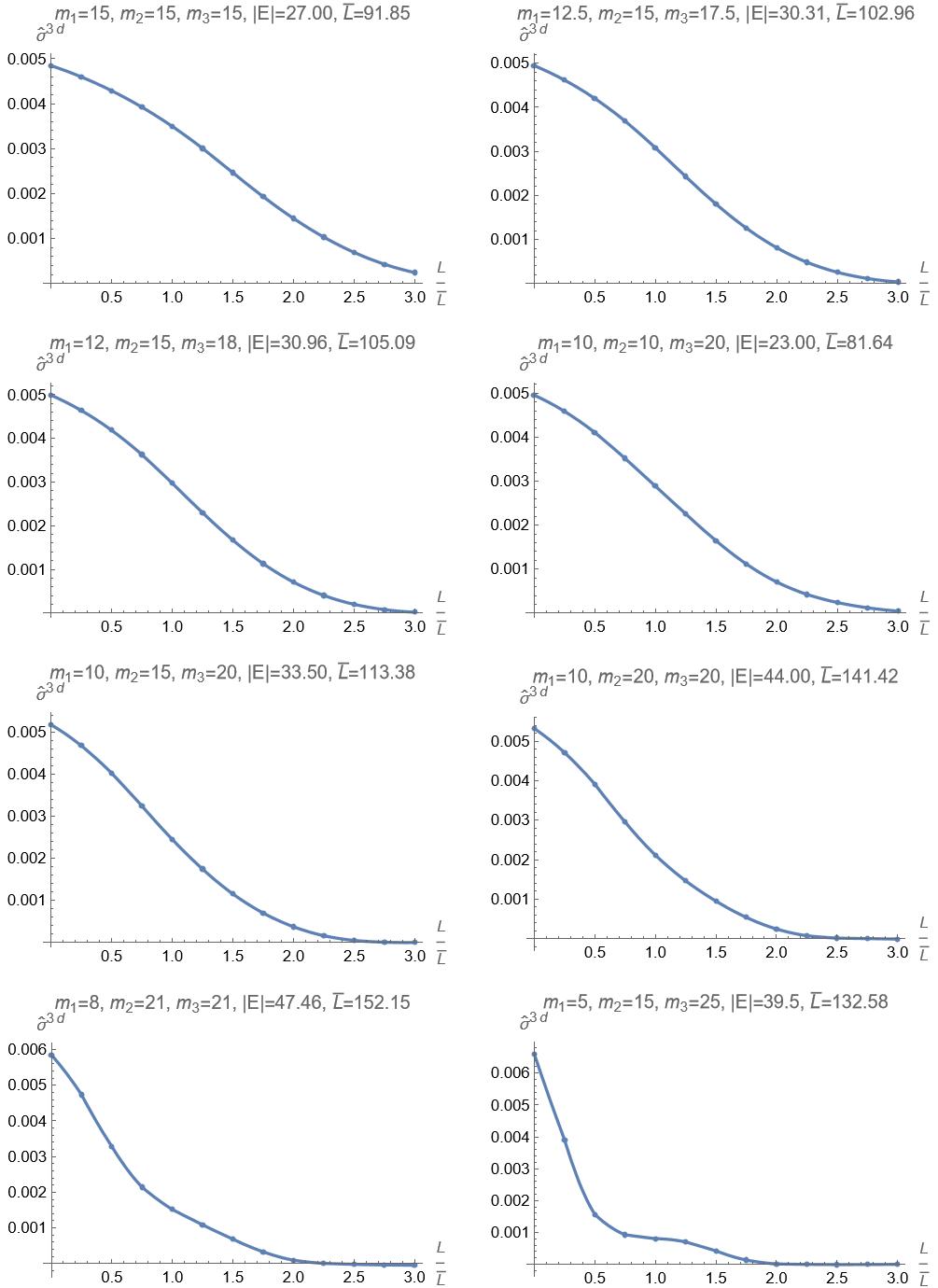}
	\caption{$L$ dependence of $\hat\sigma(E,L)$ for MKTL mass sets.}
	\label{mktl3d}
\end{figure}

In subsequent work, we plan to compare the lifetimes extracted from the simulations of \cite{simulate} to the predictions from the flux-based theory \cite{flux_based} by using the results presented in Table \ref{sb3}. In Fig. \ref{mktl3d}, we plot the $L-$dependence for these mass sets by scatterplots and interpolation curves. \footnote{A few of the observed negative values in fig \ref{mktl3d} are all quite small (in absolute values) and we interpret them as being consistent with positive 
infinitesimals ($0^+$).}  \\\\
{\bf Falloff at large $L$}

The values of regularized phase-volume for which we have reliable answers do not appear to have either power-law falloff or an exponential falloff at large values of $L$. For the regularized phase-volume in the planar case, we will see that it appears to have a power-law falloff at large $L$.

\begin{table}
\begin{center}
   \begin{tabular}{ |c|c|c|c|c| } 
 \hline
 Masses & $|E|$ & $L$ & $\bar\sigma(E,L)$ & $\hat\sigma(E,L)$\\ 
 \hline
 15,15,15 & 27.00 & 91.85 & $1.39\times10^{9}$& 0.00905\\ 
 12.5,15,17.5 & 30.31 & 102.96 & $8.13\times10^{8}$& 0.00598\\ 
 12,15,18 & 30.96 & 105.09 & $7.04\times10^{8}$& 0.00547\\
 10,10,20 & 23.00 & 81.64 & $4.05\times10^{8}$& 0.0112\\
 10,15,20 & 33.50 & 113.38 & $3.38\times10^{8}$& 0.00370\\
 10,20,20 & 44.00 & 141.42 & $2.58\times10^{8}$& 0.00161\\
 8,21,21  & 47.46 & 152.15 & $7.75\times10^{7}$& 0.000959\\
 5,15,25  & 39.5 & 132.58 & $7.82\times10^{6}$& 0.000804\\
 \hline
\end{tabular}
\caption{$\bar\sigma(E,L)$ for 8 Mass sets of MKTL \cite{simulate}} 
\label{sb3}
\end{center}
\end{table}

\section{Planar three-body problem $\bar\sigma^{2d}(E,L)$} 
\label{sec:planar}

Let us consider a special case of the three-body problem where the dynamics is restricted to a plane. This means that the initial momenta of all three bodies lie along the plane defined by the location of the three bodies. Since this case is analogous to the 3D case, we will keep the discussion short and highlight only key results. 

\subsection{Reduction of $\sigma^{2d}(E,L)$ to $S^2$}
The three-body problem is defined by the same Hamiltonian as \eqref{hm}, except now the coordinates and momenta are two-dimensional. If the motion is restricted to the $x-y$ plane of the 3D Cartesian coordinate system, then we have $L_x=L_y=0$, and $L_z(=L)$ is the only relevant component. The bare phase-volume for the planar three-body system is defined as
\begin{equation}\label{bars2} 
	\sigma^{2d}(E,L) := \int \prod_{c=1}^{3} (d^2r_c d^2p_c) \delta^{(2)}(\vec R_{cm})\delta^{(2)}(\vec P_{cm}) \delta(J-L) \delta(H-E)
\end{equation}
Its dimensions are 
\begin{equation}\label{dim2}
		\left[\sigma^{2d}(E,L)\right] = \frac{M^{3/2}}{E^{5/2}}~ \alpha^3 \equiv [\hbar]^2 T 
\end{equation} 
which is the same as the 3D case \eqref{dim3}. After performing the momenta integrations in \eqref{bars2} we get
\begin{equation}\label{smi2}
		\sigma^{2d}(E,L) = (4\pi) \left(\frac{M_3}{M}\right) \int_{T^{2d}_{\text{eff}}\geq0} ~\prod_{c=1}^{3} d^2r_c~ \delta^{(2)}(\vec R_{cm}) \sqrt{\frac{2T^{2d}_{\text{eff}}}{I}} 
\end{equation}
where
\begin{equation}
		I:= \sum_a m_a r_a^2 \quad , \quad T^{2d}_{\text{eff}} = E-V-\frac{L^2}{2I} 
\end{equation}
We perform the same sequence of coordinate change as section \ref{cc}, namely from Cartesian coordinates $x_a,y_a$ to $w,\vec R_{cm}$, and then from $w$ to $r,\theta,\phi,\psi$. After this change of coordinates (and performing the $\vec R_{cm}$ and $\psi$ integration) we get
\begin{equation}\label{sr52}
	\begin{split}
		\sigma^{2d}(E,L) = (6\sqrt{2}\pi^2) \left(\frac{M_3}{M}\right) \int \sin \theta d\theta d\phi \sqrt{\frac{1}{\overline I}}  \int dr  ~r \sqrt{-|E|r^2-{\bar V}r-\frac{L^2}{2\bar I}}^+	
	\end{split}
\end{equation}
Performing the $r$ integration (where the integration limits are set by the positivity of the integrand) gives
\begin{equation}\label{sr4}
		\sigma^{2d}(E,L) = \left(\frac{3}{8}\sqrt{2}\pi^3\right) \left(\frac{M_3}{M}\right)\frac{1}{|E|^{5/2}} \int \sin \theta d\theta d\phi \sqrt{\frac{1}{\overline I}} (-\bar V) \left(\bar V^2-\frac{2|E|L^2}{\bar I}\right)^+
\end{equation}
Hence the bare phase-volume is reduced to an integration over $S^2$ defined by $\theta,\phi$. Let us validate the result by performing $L-$integration of the expression \eqref{sr4}
\begin{equation}
	\begin{split}
		&\int dL~ \sigma^{2d}(E,L) \\
		&= \left(\frac{3}{8}\sqrt{2}\pi^3\right) \left(\frac{M_3}{M}\right)\frac{1}{|E|^{5/2}} \int \sin \theta d\theta d\phi \sqrt{\frac{1}{\overline I}} (-\bar V) \int_{-\sqrt{\frac{\bar V^2 I}{2E}}}^{\sqrt{\frac{\bar V^2 I}{2E}}} dL \left(\bar V^2-\frac{2|E|L^2}{\bar I}\right)\\
		&= \left(\frac{3}{8}\sqrt{2}\pi^3\right) \left(\frac{M_3}{M}\right)\frac{1}{|E|^{5/2}} \int \sin \theta d\theta d\phi \sqrt{\frac{1}{\overline I}} (-\bar V) \left[\frac{2\sqrt{2}}{3}\sqrt{\frac{\bar I}{|E|}}(-\bar V)^3\right]\\
		&= \frac{\pi^3}{2} \left(\frac{M_3}{M}\right)\frac{1}{|E|^{3}} \int \sin \theta d\theta d\phi~ \bar V^4
	\end{split}
\end{equation}
which matches with the expression of $\sigma^{2d}(E)$ in Paper I as expected.

\subsection{Reduction of $\sigma^{2d}_{\text{ref}',s}(E,L)$ to $S^2$}

We define the reference phase-volume by replacing $V\rightarrow V_F$ (for each body) in the expression for the bare phase-space volume and choosing the integration domain $D_s$ defined by the condition $E_B\leq E/2$. We find it convenient to implement the regularization using the simpler scheme defined by the condition $u_B\geq 1/2$, and calculating the appropriate compensator. We subtract the integrand of the reference phase-volume from the integrand of the bare phase-volume, and then perform numerical integration over $S^2$ to find the regularized phase-volume.

We derive $\sigma^{2d}_{\text{ref}'}(E,L)$ by replacing the potential $V$ by $V_{F}$ in the expression \eqref{sr52} and imposing the condition $u_B\geq 1/2$ on the integration. After performing the $r$ integration analytically, the result is 
\begin{equation}\label{sr4r}
\begin{split}
    \sigma&^{2d}_{\text{ref}',s}(E,L) = \left(\frac{3}{8}\sqrt{2}\pi^3\right) \left(\frac{M_3}{M}\right)\frac{1}{|E|^{5/2}} \int \sin \theta d\theta d\phi \sqrt{\frac{1}{\overline I}}~\theta\left(\bar V_{F}^2-\frac{2|E|L^2}{\bar I}\right) 
    \\& \times \Bigg[ \theta(r_{max}-R_+)(-\bar V_{F}) \left(\bar V_{F}^2-\frac{2|E|L^2}{\bar I}\right) \\ & +\theta\left(R_+-r_{max}\right)\theta(r_{max}-R_-)\bigg[\frac{1}{2}(-\bar V_{F}) \left(\bar V_{F}^2-\frac{2|E|L^2}{\bar I}\right) \\&- \frac{1}{\pi}(-\bar V_{F}) \left(\bar V_{F}^2-\frac{2|E|L^2}{\bar I}\right) \tan^{-1} \left(\frac{-\bar V_{F}+4\bar V_B}{2\sqrt{2\bar V_B(\bar V_{F}-2\bar V_B)-|E|L^2/(2\bar I)}}\right) \\& -\frac{16}{3\pi} \left(2\bar V_B(\bar V_{F}-2\bar V_B)-\frac{|E|L^2}{2\bar I}\right)^{3/2} \\& -\frac{2}{\pi} \bar V_{F}(\bar V_{F}-4\bar V_B)\left(2\bar V_B(\bar V_{F}-2\bar V_B)-\frac{|E|L^2}{2\bar I}\right)^{1/2}\bigg]\Bigg]
\end{split}
\end{equation}

where
\begin{equation}
	R_\pm := \frac{-\bar V_{F}\pm\sqrt{\bar V_{F}^2-\frac{2|E|L^2}{\bar I}}}{2|E|} \quad , \quad r_{max} := \frac{2\bar V_B}{|E|}
\end{equation}
Hence we have reduced the reference phase-volume to an integration over $S^2$ defined by $\theta,\phi$.

\subsection{Compensator $\Delta\sigma^{2d}(E,L)$}
Using the effective variables for hierarchical configurations, the compensator can be expressed as
\begin{equation}
	\begin{split}
		\Delta\sigma^{2d}_{s}(E,L) = & \int dE_B \int d L_B\int d^2 r_B  d^2 p_B  \delta(H_B-E_B)\delta(J_B- L_B) \times \\& \times \int d^2 r_F d^2 p_F\delta(H_F-(E-E_B))\delta( J_F-( L- L_B))
	\end{split}
\end{equation}
where the integration domain is restricted by the conditions $u_B\geq 1/2$ and $E_B\geq E/2$. After integrating over $\vec p_B,\vec p_F$ (by introducing conjugate variables), we find
\begin{equation}\label{ebm2}
	\begin{split}
		&\Delta\sigma^{2d}_{s}(E,L) = 2 \left(\frac{M_3}{M}\right)^{1/2} \int dE_B \int d L_B \int \frac{d^2 r_B d^2 r_F}{ r_B r_F}~\times \\& \times \left(E_B + \frac{\alpha_B}{r_B}-\frac{L_B^2}{2\mu_B r_B^2}\right)_+^{-1/2}  \left(E_F + \frac{\alpha_F}{r_F}-\frac{L_F^2}{2\mu_F r_F^2 }\right)_+^{-1/2}
	\end{split} 
\end{equation} 
where
\begin{equation}
	E_F=E-E_B \quad ,\quad L_F=L-L_B 
\end{equation}
Since the integrand depends only on the magnitude of the $\vec r_B$ and $\vec r_F$, the integration over their directions becomes trivial. After changing variables from $r_B,r_F$ to $u_B,u_F$ as defined in \eqref{udef}, we find
\begin{equation}
	\begin{split}
		&\Delta\sigma^{2d}_{s}(E,L) = (8\pi^2) \left(\frac{M_3}{M}\right)^{1/2} \frac{(\alpha_B\alpha_F)}{|E|^2} \int dE_B \int d L_B   \int \frac{d u_B}{u_B^2}\int \frac{d u_F}{u_F^2}~\times \\& \times  \left(E_B + u_B|E| - \frac{L_B^2|E|^2}{2\mu_B\alpha_B^2}u_B^2 \right)_+^{-\frac{1}{2}}  \left(E-E_B +u_F|E| - \frac{(L-L_B)^2|E|^2}{2\mu_F\alpha_F^2}u_F^2\right)_+^{-\frac{1}{2}}
	\end{split} 
\end{equation}
After performing the integration over $u_F$ and $u_B$ analytically, we find
\begin{equation}
    \begin{split}
        &\Delta\sigma^{2d}_{s}(E,L) 
		=(8\pi^2) \left(\frac{M_3}{M}\right)^{1/2} \frac{(\alpha_B\alpha_F)}{|E|^2}  \int_{{\cal R}_1}  dE_B~ dL_B~\times\\& \quad \quad \quad \quad \times \left[\frac{\pi}{2}\frac{|E|}{(|E|+E_B)^{3/2}}\right] 
		\left[\frac{\pi}{2}\frac{|E|}{(-E_B)^{3/2}}\right]
          \\&+ (8\pi^2) \left(\frac{M_3}{M}\right)^{1/2} \frac{(\alpha_B\alpha_F)}{|E|^2}  \int_{{\cal R}_2}  dE_B~ dL_B~ \left[\frac{\pi}{2}\frac{|E|}{(|E|+E_B)^{3/2}}\right]\times \\&\times \Bigg\{
		\Bigg[\frac{2}{E_B}\sqrt{\frac{|E|}{2}+E_B-\frac{|E|^2L_B^2}{8k_B}}+\frac{|E|}{4(E_B)^{3/2}}\log\left(\frac{-1+\frac{|E|+4E_B}{4\sqrt{E_B}\sqrt{\frac{|E|}{2}+E_B-\frac{|E|^2L_B^2}{8k_B}}}}{1+\frac{|E|+4E_B}{4\sqrt{E_B}\sqrt{\frac{|E|}{2}+E_B-\frac{|E|^2L_B^2}{8k_B}}}}\right)\Bigg]\theta(E_B)
  \\& +\Bigg[\frac{\pi}{4}\frac{|E|}{(-E_B)^{3/2}}-\frac{2}{(-E_B)}\sqrt{\frac{|E|}{2}+E_B-\frac{|E|^2L_B^2}{8k_B}}\\&-\frac{|E|}{2(-E_B)^{3/2}}\arctan\left(\frac{|E|+4E_B}{4\sqrt{(-E_B)}\sqrt{\frac{|E|}{2}+E_B-\frac{|E|^2L_B^2}{8k_B}}}\right)\Bigg]\theta(-E_B) \Bigg\}
    \end{split}
    \label{planar_compensator}
\end{equation}
where ${\cal R}_1$ and ${\cal R}_2$ are the two regions of integration given by
\begin{equation}
	\begin{split}
		{\cal R}_1 &\equiv \left(\frac{E}{2}\leq E_B\leq \frac{E}{4}\right) \text{and} \left(E_B\leq E+ \frac{k_F}{2(L-L_B)^2}\right)\text{and} \left( \frac{-k_B}{2L_B^2}\leq E_B\leq \frac{E}{2}+\frac{E^2L_B^2}{8k_B}\right)  \\
		{\cal R}_2 &\equiv \left(\frac{E}{2}\leq E_B\leq \infty\right) \text{and} \left(E_B\leq E+ \frac{k_F}{2(L-L_B)^2}\right)\text{and} \left(  \frac{E}{2}+\frac{E^2L_B^2}{8k_B}\leq E_B\right) 
	\end{split} 	
\end{equation}
The remaining integration over $L_B,E_B$ is performed numerically. The compensator satisfies the $L$ integration validation performed numerically, i.e. it is consistent with the compensator $\bar\sigma^{2d}(E)$ in Paper I.

\subsection{Evaluation of $\bar\sigma^{2d}(E,L)$}

{\bf Numerical integration and addition of Compensator}

The bare phase-volume $\sigma^{2d}(E,L)$ has three coincidence singularities which lie on the $\theta=\pi/2$ equator at $\phi=0,2\pi/3,4\pi/3$. The integrand of the regularized phase-volume turns out to be (incidentally) regular at the location of the three coincidence singularities, leading to a finite value of integration. By numerical integration (using Mathematica), we find the regularized phase-volume for the $u_B\geq1/2$ scheme. We perform numerical integration with an excision around coincidence singularities and take this excision to zero. We add the compensator to derive the answers for the desired $E_B\leq E/2$ scheme.
\\\\
{\bf Normalization}

We define the normalized and dimensionless phase-volume $\hat\sigma^{2d}(E,L)$ as
\begin{equation}
    \begin{split}
        \hat\sigma^{2d}(E,L) &:= \frac{1}{(2\pi)^4}\frac{M}{M_3}\frac{|E|^3}{\sigma_0^{2d}} (\bar L^2)^{1/2} \bar\sigma^{2d}(E,L) \\ \sigma_0^{2d} &:= \sum_{s=1}^3 (\alpha_{B,s}\alpha_{F,s})^2  \quad \text{and} \quad \bar L^2 := \frac{1}{2|E|}\frac{\sum_{s=1}^3 k_s}{3} 
    \end{split}
\end{equation}
This normalization is a simple modification of the normalization used in Paper I. \\\\
{\bf $L$ dependence and $L$ integration check}

\begin{figure}
	\includegraphics[scale=1]{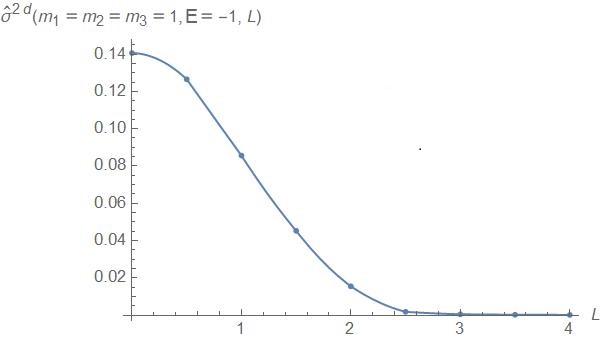}
	\caption{$L$ dependence of $\hat\sigma^{2d}(E,L)$, fixing $m_1=m_2=m_3=1$ and $|E|=1$}
	\label{svl}
\end{figure}

We observe the $L$ dependence using scatterplot (and an interpolation curve) of $\hat\sigma^{2d}(E,L)$, by setting $m_1=m_2=m_3=1$, $|E|=1$. We overlap the scatterplot and interpolation curve for $\hat\sigma^{2d}(E,L)$ in figure \ref{svl}. For the values of $L$ considered, we observe that $\hat\sigma^{2d}(E,L)$ is a monotonic function of $L$, remains positive, and drops sharply as $L$ increases. The $L$ integration validation was performed numerically for $\bar\sigma(E,L)$ using the interpolation function, and it was found to be satisfied within 1\% relative error. \\\\
{\bf $\bar\sigma^{2d}(E,L)$ for 8 Mass sets of MKTL \cite{simulate}}

\begin{figure}
	\includegraphics[scale=0.6]{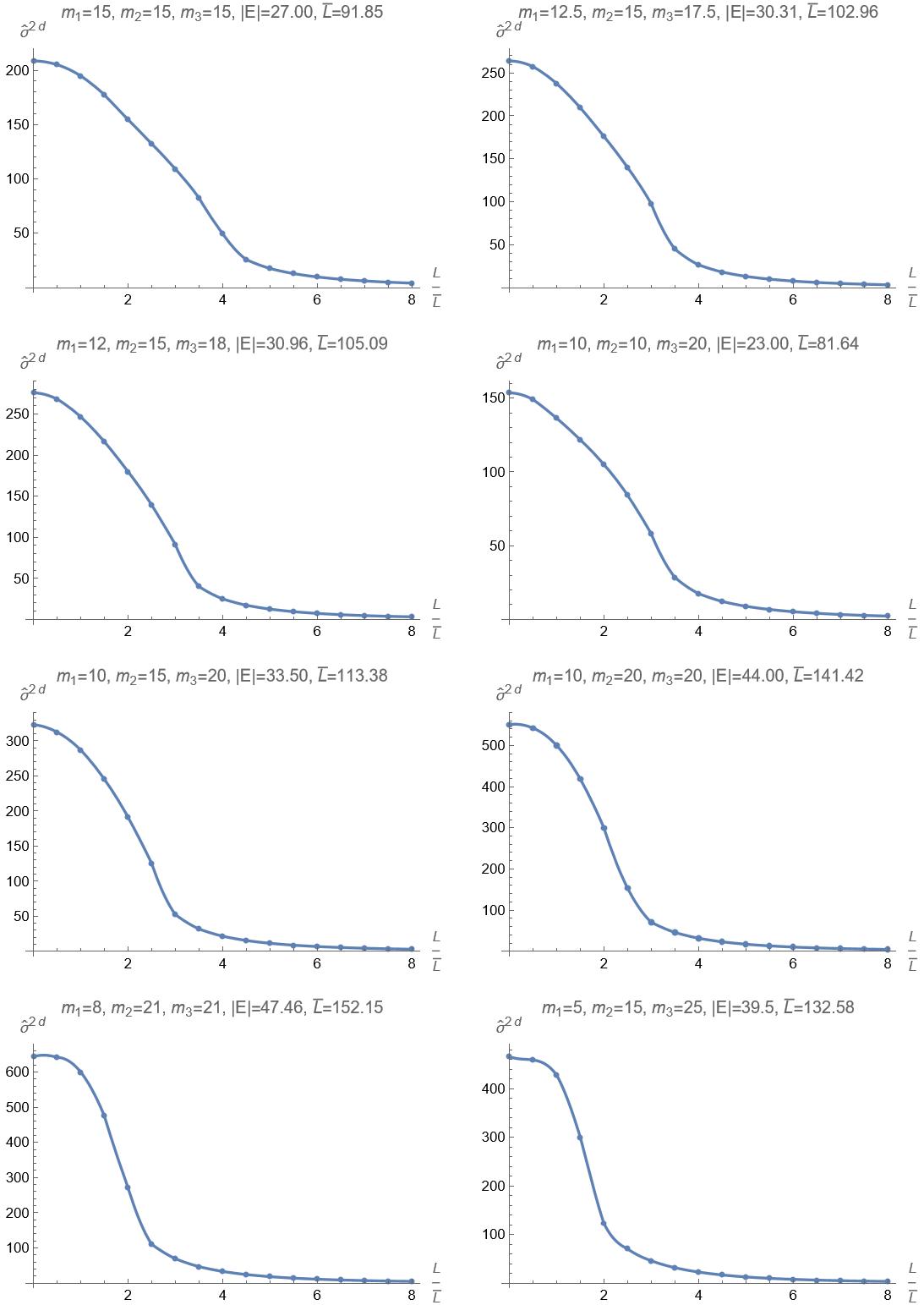}
	\caption{$L$ dependence of $\hat\sigma^{2d}(E,L)$ for MKTL mass sets.}
	\label{mktl2d}
\end{figure}

As in the 3D case, we consider the 8 mass sets of \cite{simulate} and present the results for the regularized phase-volume for these mass sets in table \ref{sb2}, and plot the $L-$dependence in Fig. \ref{mktl2d}.\\\\
{\bf Power law at large $L$}

 The values of regularized phase-volume in the planar case for which we have reliable answers appear to satisfy a phenomenological power-law at large values of $L$. For the data shown in fig \ref{svl}, at large values of $L$ we find $\hat\sigma^{2d}(E,L)\sim 1/L^{3.38}$. For the data shown in fig \ref{mktl2d}, at large values of $L$ we find $\hat\sigma^{2d}(E,L)\sim 1/L^{a}$, where $a=3.08, 2.95, 2.93, 2.90, 2.86, 2.83, 2.78, 2.76$ sequentially for each mass set. These power laws should be tested at even higher values of $L$ by gathering higher precision data.   

\begin{table}
\begin{center}
   \begin{tabular}{ |c|c|c|c|c| } 
 \hline
 Masses & $|E|$ & $L$ & $\bar\sigma^{2d}(E,L)$ & $\hat\sigma^{2d}(E,L)$\\ 
 \hline
 15,15,15 & 27.00 & 91.85 & $4.24\times10^{11}$ & $1.95\times10^{2}$\\ 
 12.5,15,17.5 & 30.31 & 102.96 & $3.57\times10^{11}$ & $2.38\times10^{2}$\\ 
 12,15,18 & 30.96 & 105.09 & $3.38\times10^{11}$& $2.47\times10^{2}$\\
 10,10,20 & 23.00 & 81.64 & $1.18\times10^{11}$& $1.36\times10^{2}$\\
 10,15,20 & 33.50 & 113.38 & $2.56\times10^{11}$& $2.87\times10^{2}$\\
 10,20,20 & 44.00 & 141.42 & $4.43\times10^{11}$& $5.00\times10^{2}$\\
 8,21,21  & 47.46 & 152.15 & $2.90\times10^{11}$& $6.00\times10^{2}$\\
 5,15,25  & 39.5 & 132.58 & $5.97\times10^{10}$& $4.28\times10^{2}$\\
 \hline
\end{tabular}
\caption{$\bar\sigma^{2d}(E,L)$ for 8 Mass sets of MKTL \cite{simulate}} 
\label{sb2}
\end{center}
\end{table}

\section{Summary and discussion}
\label{sec:disc}

The main results of this paper are the development of an evaluation method for the regularized phase-volume $\bsig(E,L;m_a)$ and the achievement of first evaluations of this quantity.

\vspace {0.5cm}  \noindent {\bf Development}. We find that analytic integration is possible over both the configuration scale and the inclination angle between the triangle plane and $\vec{L}$. This was done both for the bare integrand in \eqref{frs} and 
the reference' integrand in (\ref{rwe},\ref{ref_reduced_schem2},\ref{qgz},\ref{qlz}), 
reducing both to a numerical integral over $S^3$, the 3-sphere. The difference integrand becomes regular, and so the integral is convergent.
 We find reliable answers for the regularized phase-volume by considering excisions at the locations of the three coincidence singularities and taking the limit where the excision size vanishes. The compensator was reduced to a 3d numerical integral in \eqref{compensator_reduced}. 

Comparing with the $\bsig(E)$ case in Paper I, where the numerical integration is over $S^2$, we note an increase in the dimension of the integration domain which is a result of the reduced symmetry. Also, in Paper I the compensator was determined analytically thanks to an involution symmetry. $\bsig(E,L)$ loses this symmetry ,leading to increased computation. On a personal computer, evaluating $\bsig(E,L)$ typically takes between 30 minutes and an hour, whereas evaluating $\bsig(E)$ requires only a few minutes.

In the planar case, the bare and reference phase-volumes were reduced to a numerical integration over $S^2$ in (\ref{sr4}, \ref{sr4r}), respectively. 
The compensator was reduced to a numerical integration over a 2d domain in \eqref{planar_compensator}.  
 Comparing with the $\bsig(E)$ case in Paper I, we find that the integration domain for in the ref' scheme remains unchanged, while the determination of the compensator was previously achieved analytically, and this is lost due to the loss of the involution symmetry. 

\vspace {0.5cm} \noindent {\bf Evaluation}. The dependence of $\bsig(E,L;m_a)$ on one of its parameters is determined by dimensional analysis, while the dependence on the rest is through a numerical function, which can only be evaluated and presented at discrete values of the parameters. We evaluate them for the mass sets considered in \cite{simulate}: $L$-dependence graphs are shown in Fig. \ref{svl3} and \ref{mktl3d}, 
and Table \ref{sb3} 
presents $\bsig$ values for the specific $L$-values of \cite{simulate}. In the planar case, we present the $L$-dependence in Fig. \ref{svl} 
and \ref{mktl2d} 
and for specific $L$ values, evaluations are presented in Table \ref{sb2}. 

\vspace {0.5cm} \noindent {\bf Validation}.  Our first type of validation is through the $L$-integration test: since the various quantities (regularized values in reference' scheme and the compensator) are already known for $\bsig(E)$, the distributed quantities $\bsig(E,L)$ must integrate over $dL$ to the former. We find that our evaluations satisfy this constraint, and consider it a strong validation of our method and evaluation.

The second type of validation is qualitative and concerns the positiveness of $\bsig(E,L)$. Since these are regularized values, evaluated by the subtraction of two positive integrands, there is no guarantee {\it a priori} that the difference would be positive. We consider the positiveness of the $\bsig(E,L)$ evaluations to be a positive sign regarding the definition of the regularization scheme and the correctness of the evaluation. \footnote{Note that for non-comparable masses, we have discovered already in paper I that the regularized phase-volume could turn negative, and we interpreted that as a signal for the breakdown of the statistical theory, which is not surprising.}

\vspace {0.5cm} \noindent {\bf Analysis of results}. We note that for high values of $L$,  $\bsig(E,L)$ approaches zero fast (while remaining positive), see Fig. (\ref{svl3}--\ref{mktl2d}). In the 3d case, we were unable to fit the high $L$ behavior with an analytic expression, while in the 2d case, preliminary analysis indicates a power-law decay with index in the range of 2.8--3.4.

\vspace {0.5cm} \noindent {\bf Open questions}. It would be interesting to analyze simulation data (such as \cite{calE}) to extract lifetimes and compare with flux-based predictions that rely on the results of this paper. 

It would be interesting to determine the large $L$ behavior and to explain it by the theory. 

\subsection*{Acknowledgments}

BK acknowledges support by the Israel Science Foundation (grant No. 1345/21) and by a grant from Israel's Council of Higher Education. YD is supported by the Israel Science Foundation (ISF) grant No. 1698/22. We would like to thank Ofek Birnholtz for useful discussions.

\appendix

\section{Analytic reduction of $\bar\sigma(E,L)$}
\label{app:A}

\subsection{Derivation of expression \eqref{frs} for $\sigma(E,L)$ \label{apa}}

Let us begin with expression \eqref{smi}. We perform the planar reduction \eqref{plr}, change variables from Cartesian coordinates to $w,\vec r_{cm}$, perform the integration over $\vec r_{cm}$, and change variables from $w$ to $r,\theta,\phi,\psi$. By doing this, we get
\begin{equation}\label{sr6}
	\begin{split}
		\sigma(E,L) = &3\sqrt{2} \pi \left(\frac{M_3}{M}\right)^{3/2} \int \sin \theta d\theta d\phi d\psi \int \sin \theta_n d\theta_n d\phi_n dr \\ &\times\frac{\bar A_\Delta}{\sqrt{\overline{\det I}}} ~r~ \sqrt{-|E|r^2-{\bar V}r-\frac{1}{2}{\bar I}^{-1}_{ij}L^iL^j}^+	
	\end{split}
\end{equation}
Choosing the orientation of $\vec L$ along the $\theta_n=0$ makes the integrand of \eqref{sr6} independent of $\phi$, and hence its integration produces a $2\pi$ factor. Using the following identity for the three-body system
\begin{equation}
	\begin{split}
		\det I^{(2)} = \frac{M_3}{M} (2A_\Delta)^2 
	\end{split}
\end{equation}
we get
\begin{equation}\label{exc}
	\frac{\bar A_\Delta}{\sqrt{\overline{\det I}}} = \frac{1}{2}\left(\frac{M}{M_3}\right)^{1/2}\sqrt{\frac{1}{\overline I}}
\end{equation}
Substituting \eqref{exc} in \eqref{sr6} we get
\begin{equation}\label{sr5}
	\begin{split}
		\sigma(E,L) = &3\sqrt{2} \pi^2 \left(\frac{M_3}{M}\right) \int \sin \theta d\theta d\phi d\psi \sqrt{\frac{1}{\overline I}} \times \\& \times\int \sin \theta_n d\theta_n \int dr  ~r \sqrt{-|E|r^2-{\bar V}r-\frac{1}{2}{\bar I}^{-1}_{ij}L^iL^j}^+	
	\end{split}
\end{equation}
Performing the $r$ integration with the limits set by the positivity of the integrand gives
\begin{equation}\label{sr43}
	\begin{split}
		\sigma(E,L) = &\frac{3\sqrt{2}}{16} \pi^3 \left(\frac{M_3}{M}\right)\frac{1}{|E|^{5/2}} \int \sin \theta d\theta d\phi d\psi \sqrt{\frac{1}{\overline I}}\times \\& \times(-\bar V) \int \sin \theta_n d\theta_n  \left(\bar V^2-2|E|{\bar I}^{-1}_{ij}L^iL^j\right)^+
	\end{split}
\end{equation}
with
\begin{equation}\label{iic}
	\begin{split}
		{\bar I}^{-1}_{ij}L^iL^j = \frac{L^2}{\det \bar I^{(2)}} \bigg( &\bar I^{yy} \sin^2 \theta_n \cos^2 \psi+\bar I^{xx}\sin^2 \theta_n \sin^2 \psi \\&+ 2\bar I^{xy}\sin^2 \theta_n \cos\psi\sin\psi + \frac{\det \bar I^{(2)}}{\bar I} \cos^2 \theta_n \bigg)
	\end{split}
\end{equation}
where $\bar I^{ij}$ are functions of $\theta,\phi$. Changing the variable $\theta_n$ to $\tau:=\cos\theta_n$,
the expression \eqref{sr43} becomes
\begin{equation}\label{sr4m}
	\begin{split}
		\sigma(E,L) = \frac{3\sqrt{2}}{16} \pi^3 \left(\frac{M_3}{M}\right)\frac{1}{|E|^{5/2}} \int \sin \theta d\theta d\phi d\psi \sqrt{\frac{1}{\overline I}} (-\bar V) \int d\tau \left(A+B\tau^2\right)^+
	\end{split}
\end{equation}
where
\begin{equation}\label{abdef}
		\begin{split}
			\frac{A}{2|E|L^2} &:= A_1 - A_2 \quad,\quad \frac{B}{2|E|L^2} := A_2 - B_1 \\ A_1 &= \frac{\bar V^2}{2|E|L^2} \quad ,\quad A_2=\frac{\frac{1}{2}\left(\overline{I}+\overline{I}(\psi)\right)}{\overline{\det I^{(2)}}} \quad,\quad B_1 = \frac{1}{\overline I} \\
			I(\psi) &:= (I^{yy}-I^{xx})\cos 2\psi + 2I^{xy}\sin 2\psi 
	\end{split} 
\end{equation}
where $(x)^+:={\rm max} \{x, 0 \}$ denotes the ramp function.

We have $A_1\geq0$ and $B_1\geq0$. It can be shown that $I(\psi)^2\leq I^2 - 4\det I^{(2)}$. Using this, it can be shown that $A_2\geq0$ and $A_2\geq B_1$. The $\tau$ integration in \eqref{sr4m} is performed as 
\begin{equation}\label{pint}
	\begin{split}
		\int d\tau \left(A+B\tau^2\right)^+ &= \theta(A) \left(\int_{-1}^{+1}d\tau\left(A+B\tau^2\right)\right) \\&\quad + \theta(-A)\theta(A+B)\left(\int_{-1}^{+1}d\tau\left(A+B\tau^2\right) -\int_{-\sqrt{|A/B|}}^{+\sqrt{|A/B|}}d\tau\left(A+B\tau^2\right) \right)
		\\&= \theta(A+B) \left(\int_{-1}^{+1}d\tau\left(A+B\tau^2\right) - \theta(-A)\int_{-\sqrt{|A/B|}}^{+\sqrt{|A/B|}}d\tau\left(A+B\tau^2\right) \right) \\&  = \theta(A+B) \left[\left(2A+\frac{2}{3}B\right)-\theta(-A)\frac{4}{3}A\sqrt{\bigg|\frac{A}{B}\bigg|}~ \right]
	\end{split}
\end{equation}
Substituting \eqref{pint} in \eqref{sr4m}, we get the expression \eqref{frs} for $\sigma(E,L)$ 
\begin{equation}\label{si3}
\begin{split}
    \sigma(E,L) = &\frac{\pi^3}{4\sqrt{2}}  \left(\frac{M_3}{M}\right)\frac{1}{|E|^{5/2}}\int \sin \theta~ d\theta~ d\phi~ d\psi~ \times \\& \times\sqrt{\frac{1}{\overline{I}}}(-\bar V)\left[ (3A+B) - 2A \sqrt{\bigg|\frac{A}{B}\bigg|}\theta(-A) \right]\theta(A+B)
\end{split}
\end{equation} 
Let us check the positivity of the integrand in \eqref{si3}. We have the result $B\geq0$. First, if $A\geq0$, the integrand becomes $(3A+B)$ times a positive factor, which means the integrand is positive definite. Second, if $A\leq0$, the factor $\theta(A+B)$ in the integrand becomes relevant.
In this case, the integrand can be expressed as $(2x^{3/2}-3x+1)$ times a positive factor, where $x:=\frac{-A}{B}$ and $0\leq x\leq 1$. It can be checked that $(2x^{3/2}-3x+1)\geq 0$ for $0\leq x\leq 1$. This shows that the integrand is positive definite for the whole integration domain.

\subsection{Writing the integrand of $\sigma(E,L)$ in $\theta,\phi,\psi$ coordinates}\label{rwi}

  We derive expressions for the factors $I, \det I^{(2)}, I(\psi)$ in $r,\theta,\phi,\psi$ coordinates. Consider the following decomposition of $w$ variable
\begin{equation}\label{wdec}
	w = w_0 + i w_i + j w_j + ij w_{ij}
\end{equation}
Inverting the relation \eqref{wdef} (along with the center of mass constraint) in terms of the \eqref{wdec}, we find
\begin{equation}\label{ciw}
	\begin{split}
		x_3 &= \frac{1}{-\frac{m_3}{m_1}+\frac{\sqrt{3}-1}{2}}\left(-\frac{\sqrt{3}/2}{\frac{m_2}{m_1}+\frac{1}{2}}w_0-w_j\right) ~ , ~ x_2 = \frac{2}{\sqrt{3}}w_j + x_3 ~,~ x_1 = w_0+\frac{x_2+x_3}{2}	 \\
		y_3 &= \frac{1}{-\frac{m_3}{m_1}+\frac{\sqrt{3}-1}{2}}\left(-\frac{\sqrt{3}/2}{\frac{m_2}{m_1}+\frac{1}{2}}w_i-w_{ij}\right) ~,~ y_2 = \frac{2}{\sqrt{3}}w_{ij} + y_3 ~,~ y_1 = w_i+\frac{y_2+y_3}{2}
	\end{split} 
\end{equation}
So, the Cartesian components of $\vec r_{12},\vec r_{13},\vec r_{23}$ are expressed as
\begin{equation}\label{rcom}
	\begin{split}
		x_1-x_2 &=w_0-\frac{1}{\sqrt{3}}w_j ~~,~~ x_1-x_3 = w_0+\frac{1}{\sqrt{3}}w_j ~~,~~ x_2-x_3=\frac{2}{\sqrt{3}}w_j \\
		y_1-y_2 &=w_i-\frac{1}{\sqrt{3}}w_{ij} ~~,~~ y_1-y_3 = w_i+\frac{1}{\sqrt{3}}w_{ij} ~~,~~ y_2-y_3=\frac{2}{\sqrt{3}}w_{ij}
	\end{split} 
\end{equation}
These Cartesian components are expressed in terms of $r,\theta,\phi,\psi$ coordinates using \eqref{spdef}. This enables us to write the expressions for the necessary quantities in $r,\theta,\phi,\psi$ coordinates. First, $\det I^{(2)}$ can be written in terms of the area ($A$) of the triangle defined by the three bodies
\begin{equation}
    \det I^{(2)} = \frac{M_3}{M} (2A)^2
\end{equation}
where
\begin{equation}
	A = \frac{1}{4} \sqrt{(r_{12}^2+r_{23}^2+r_{31}^2)^2-2(r_{12}^4+r_{23}^4+r_{31}^4)}
\end{equation}
and so it becomes
\begin{equation}
		\det I^{(2)} = r^4\frac{3}{4}\frac{M_3}{M}\cos^2\theta 
\end{equation} 

The moment of inertia, $I$, can be written as
\begin{equation}
	I = \frac{m_1 m_2}{M} r_{12}^2 + \frac{m_2 m_3}{M} r_{23}^2 + \frac{m_1 m_3}{M} r_{13}^2 
\end{equation}
and so it becomes
\begin{equation}
		\begin{split}
			I = \frac{r^2}{M} \bigg[&m_1m_2\left(1-\sin\theta\cos\left(\phi-\frac{2\pi}{3}\right)\right)+m_1m_3\left(1-\sin\theta\cos\left(\phi+\frac{2\pi}{3}\right)\right)\\&+m_2m_3\left(1-\sin\theta\cos\left(\phi\right)\right)\bigg]
	\end{split}	
\end{equation}

For the expression $I(\psi)=(I^{yy}-I^{xx})\cos2\psi+2I^{xy}\sin2\psi$ we find
\begin{equation}\label{icom}
	\begin{split}
			I^{yy}-I^{xx}=\frac{r^2}{2M}\bigg[&(m_1m_2+m_1m_3-2m_2m_3)\left(\cos^2\frac{\theta}{2}+\cos2\phi\sin^2\frac{\theta}{2}\right)\\&+2(m_1m_2+m_1m_3+m_2m_3)\cos\phi\sin\theta\\&-\sqrt{3}m_1(m_2-m_3)\sin2\phi\sin^2\frac{\theta}{2} \bigg] \\	I^{xy}=\frac{r^2}{4M}\bigg[&-\sqrt{3}m_1(m_2-m_3)\left(\cos^2\frac{\theta}{2}-\cos2\phi\sin^2\frac{\theta}{2}\right)\\&+2(m_1m_2+m_1m_3+m_2m_3)\sin\phi\sin\theta\\&+(m_1m_2+m_1m_3-2m_2m_3)\sin2\phi\sin^2\frac{\theta}{2} \bigg] 
	\end{split} 
\end{equation}

\subsection{$L-$integration validation for the expression \eqref{frs} }\label{liv3}

We validate the correctness of \eqref{frs} using the expectation that 
\begin{equation}
	\int d^3L ~\sigma(E,L) = \sigma(E)
\end{equation}
where $\sigma(E)$ is the bare phase-volume evaluated in Paper I.
\begin{equation}\label{lid}
\begin{split}
    \text{LHS} = \frac{\pi^3}{4\sqrt{2}}  &\left(\frac{M_3}{M}\right)\frac{1}{|E|^{5/2}}\int \sin \theta~ d\theta~ d\phi~ d\psi~ \sqrt{\frac{1}{\overline{I}}}(-\bar V)\times \\& \times\int d^3L \left[ (3A+B) - 2A \sqrt{\bigg|\frac{A}{B}\bigg|}\theta(-A) \right]\theta(A+B) 
\end{split}
\end{equation} 
We can write
\begin{equation}
	\int d^3L = 4\pi\int dL~ L^2 
\end{equation}
due to the form of integrand. So we get
\begin{equation}\label{lin}
	\begin{split}
		&\int d^3L \left[ (3A+B) - 2A \sqrt{\bigg|\frac{A}{B}\bigg|}\theta(-A) \right]\theta(A+B) \\
		&= 4\pi \int dL~ L^2 \left[ (3A+B) - 2A \sqrt{\bigg|\frac{A}{B}\bigg|}\theta(-A) \right]\theta(A+B) \\
		&= 4\pi \int_{0}^{L_u} dL~ L^2  (3A+B) - 4\pi\int_{L_l}^{L_u} dL~ L^2~ 2A \sqrt{\bigg|\frac{A}{B}\bigg|}\\
		&= \frac{8\pi}{5} \sqrt{\bar I} \bar V^5 \frac{\overline{\det I^{(2)}}}{\frac{1}{2}\left(\overline{I}+\overline{I}(\psi)\right)}\frac{1}{2|E|}
	\end{split}
\end{equation}
where 
\begin{equation}
	L_u= \sqrt{\frac{\bar V^2\bar I}{2|E|}} \quad L_l =  \sqrt{\frac{\overline{\det I^{(2)}}}{\frac{1}{2}\left(\overline{I}+\overline{I}(\psi)\right)}\frac{\bar V^2}{2|E|}}
\end{equation}
Substitute the result \eqref{lin} in \eqref{lid}. Using the expression of $I(\psi)$ given in \eqref{abdef}, we perform the required $\psi$ integration
\begin{equation}
	\begin{split}
		\int_0^{2\pi} d\psi~ \frac{1}{\overline{I}+\overline{I}(\psi)} &= (2\pi) \left[\bar I^2-(2\bar I_{xy})^2-(\bar I_{yy}-\bar I_{xx})^2\right]^{-1/2} \\
		&= \frac{2\pi}{\sqrt{3}|\cos \theta|}\sqrt{\frac{M}{M_3}}	
	\end{split}
\end{equation}
where in the second line, we used the expressions \eqref{icom}.
In conclusion, we find
\begin{equation}
	\int d^3L ~\sigma(E,L) = \frac{\sqrt{3}}{20} \pi^5 \left(\frac{M_3}{M}\right)^{3/2} \frac{1}{|E|^4} \int~ d\theta d\phi~ \sin |2\theta|\bar V^6
\end{equation}
which matches the expression of $\sigma(E)$ in Paper I.
Hence the expression \eqref{frs} is validated by the $L$ integration test.

\bibliographystyle{unsrt}

\end{document}